\documentclass[preprint,aps,12pt,showpacs,nofootinbib,tightenlines]{revtex4}
\usepackage{amsmath}
\usepackage{amssymb}
\usepackage{epsfig}
\usepackage{graphicx}
\begin{document}
%%%%%%%%%%%%%%%%%%%%%%%%%%%%%%%%%%%%%%%%%%%%%
\newcommand{\beq}{\begin{eqnarray}}
\newcommand{\eeq}{\end{eqnarray}}
\newcommand{\non}{\nonumber\\ }

\newcommand{\calm}{{\cal M} }
\newcommand{\cala}{{\cal A} }
\newcommand{\calh}{{\cal H} }
\newcommand{\ov  }{\overline  }
\newcommand{\mw}{m_W }

\newcommand{\acp}{{\cal A}_{CP}}
\newcommand{\etap}{\eta^{(\prime)} }
\newcommand{\etar}{\eta^\prime  }

\newcommand{\pb}{\phi_B}
\newcommand{\pva}{\phi_{V}}
\newcommand{\pea}{\phi_{\eta}^A}
\newcommand{\pepa}{\phi_{\eta'}^A}
\newcommand{\pvp}{\phi_{V}^s}
\newcommand{\pep}{\phi_{\eta}^P}
\newcommand{\pepp}{\phi_{\eta'}^P}
\newcommand{\pvt}{\phi_{V}^t}
\newcommand{\pet}{\phi_{\eta}^T}
\newcommand{\pept}{\phi_{\eta'}^T}
\newcommand{\fb}{f_B }
\newcommand{\fk}{f_{\pi} }
\newcommand{\fe}{f_{\eta} }
\newcommand{\feq}{f_{\eta}^q }
\newcommand{\fes}{f_{\eta}^s }
\newcommand{\fep}{f_{\eta'} }
\newcommand{\rv}{r_{V} }
\newcommand{\re}{r_{\eta} }
\newcommand{\rep}{r_{\eta'} }
\newcommand{\mb}{m_B }
\newcommand{\xeba}{\bar{x}_2}
\newcommand{\xsba}{\bar{x}_3}
\newcommand{\res}{r_{\eta_{s\bar{s}}}}
\newcommand{\red}{r_{\eta_{d\bar{d}}}}
\newcommand{\peas}{\phi^A_{\eta_s} }
\newcommand{\peps}{\phi^P_{\eta_s}}
\newcommand{\pets}{\phi^T_{\eta_s}}
\newcommand{\pead}{\phi^A_{\eta_q}}
\newcommand{\pepd}{\phi^P_{\eta_q} }
\newcommand{\petd}{\phi^T_{\eta_q}}

\newcommand{\esl}{ \epsilon \hspace{-2.1truemm}/ }
\newcommand{\pvsl}{ p \hspace{-2.0truemm}/_V }
\newcommand{\psl}{ P \hspace{-2.4truemm}/ }
\newcommand{\nsl}{ n \hspace{-2.2truemm}/ }
\newcommand{\vsl}{ v \hspace{-2.2truemm}/ }
\newcommand{\epsl}{\epsilon \hspace{-1.8truemm}/\,  }
\newcommand{\bfkk}{{\bf k} }

%%---------------------------------------------------------

\def \epjc{ Eur. Phys. J. C }
\def \jpg{  J. Phys. G }
\def \npb{  Nucl. Phys. B }
\def \plb{  Phys. Lett. B }
\def \pr{  Phys. Rep. }
\def \prd{  Phys. Rev. D }
\def \prl{  Phys. Rev. Lett.  }
\def \zpc{  Z. Phys. C  }
\def \jhep{ J. High Energy Phys.  }

%%%%%%%%%%%%%%%%%%%%%%%%%%%%%%%%%%%%%%%%%%%%%%%%%%%%
%%
\title{$B \to \rho(\omega, \phi) \etap $ Decays and NLO contributions in the pQCD
Approach }
\author{Zhi-Qing Zhang$^{a}$ and Zhen-jun Xiao$^{a,b}$
\footnote{Electronic address: xiaozhenjun@njnu.edu.cn}}
\affiliation{{\it a. Department of Physics and Institute of
Theoretical Physics,
Nanjing Normal University, Nanjing, Jiangsu 210097, P.R.China}}
\affiliation{{\it b. Kavli Institute for Theoretical Physics China,
CAS, Beijing, 100190, China}}
\date{\today}
\begin{abstract}
By employing the perturbative QCD (pQCD) factorization approach, we
calculated some important next-to-leading-order(NLO) contributions to
the two-body charmless hadronic decays $B^+ \to \rho^+ \eta^{(\prime)}$ and
$B^0 \to \rho^0(\omega, \phi) \eta^{(\prime)}$, induced by the vertex QCD corrections,
the quark-loops as well as the chromo-magnetic penguins.
From the numerical results and phenomenological analysis we find that (a)
for $B^\pm \to \rho^\pm \etap$ decays, the partial NLO contributions to branching ratios
are small in magnitude; (b) for
$B^0 \to \rho^0(\omega,\phi) \etap$ decays, the NLO contributions can provide
significant enhancements to the leading order predictions of their branching ratios;
and (c) the pQCD predictions for the CP-violating asymmetries
$\acp^{dir}(B^\pm \to \rho^\pm \etap)$ are consistent with the data,
while the predicted $\acp(B^0 \to \rho^0(\omega)\etap)$ are generally large
in magnitude and could be tested by the forthcoming LHCb experiments.
\end{abstract}

\vspace{1cm}

%%\keywords{Keywords: Perturbative QCD, B meson decays, Branching ratio,
%%CP-violating asymmetry.}

\maketitle

\section{Introduction}

During the past decade, the B factory experiments have achieved great successes.
More than one billion events of $B \ov {B}$ production and decays have been accumulated
and analyzed by BaBar and Belle Collaborations. The forthcoming LHC experiments will
provide 2-3 orders more B meson events than the B factory, and high precision
measurements for the branching ratios and CP-violating asymmetries of many B meson rare
decays will become true within the following three to five years.
Now it becomes a very important and urgent task to reduce the uncertainty
of the theoretical predictions, in order
to test the standard mode (SM) and to find signals or evidence of
the new physics beyond the SM through the B meson experiments \cite{cpv}.

For the charmless decays $B \to M_1 M_2$ ( here $M_i$ are light mesons composed of the
light $u,d,s$ quarks), the dominant theoretical error comes from the large
uncertainty in evaluating the hadronic matrix elements $\langle M_1 M_2|O_i|B\rangle$.
In order to increase the accuracy of the SM predictions, various factorization approaches have been
proposed in recent years. The perturbative QCD (pQCD) factorization approach \cite{li2003}
, together with the so-called QCD Factorization (QCDF) \cite{bbns99} and the SCET \cite{scet1}
, are the most popular factorization approaches \cite{bbns99,scet1} being used currently
to calculate the hadronic matrix elements
\cite{bbns99,du03,kls01,luy01,li01,li05,liu06,guodq07,guo07,bs07,bs07b}.

When compared with the QCDF or SCET factorization approaches,
the pQCD approach has the following three special features:
(a) since the $k_T$ factorization is employed here,
the resultant Sudakov factor as well as the threshold resummation
can enable us to regulate the end-point singularities effectively;
(b) the form factors for $B \to M$ transition can be calculated perturbatively, although
some controversies still exist about this point;
and (c) the annihilation diagrams are calculable and play an important role in producing
CP violation.

Up to now, almost all two-body charmless $B/B_s \to M_1 M_2$ decays have been
calculated by using the pQCD approach at the leading order (LO)
\cite{kls01,luy01,li01,li05,liu06,guodq07,guo07,bs07,bs07b}.
Very recently, some next-to-leading (NLO)
contributions to $B \to K\pi$ and several $B \to PV$ decay modes \cite{nlo05,nlopv}
have been calculated, where the
Wilson coefficients at NLO accuracy are used, and the contributions from
the vertex corrections, the quark loops and the chromo-magnetic penguin operator $O_{8g}$
have been taken into account. As generally expected, the inclusion of NLO contributions
should improve the reliability of the pQCD predictions.

In previous papers \cite{liu06,guodq07}, the authors calculated the branching ratios
and CP violating asymmetries of the $B \to \rho(\omega,\phi)\etap$ decays by employing the pQCD
approach at the leading order.
Following the procedure of Ref.~\cite{nlo05}, we here would like to calculate the
NLO contributions to the  $B \to \rho(\omega, \phi)\etap$ decays by employing the low energy
effective Hamiltonian and the pQCD approach.

The remainder of the paper is organized as follows. In Sec.II, we
give a brief discussion about pQCD factorization approach. In
Sec.~III, we calculate analytically the relevant Feynman diagrams and
present the various decay amplitudes for the studied decay modes in
leading-order. In Sec.~IV, the NLO contributions from the vertex corrections, the
quark loops and the chromo-magnetic penguin amplitudes are evaluated. We show the
numerical results for the branching ratios and  CP asymmetries of $B
\to \rho(\omega,\phi)\etap$ decays in Sec.~V. The summary and some discussions are
included in the final section.

%%%%%%%%%%%%%%%%%%%%%%%%%%%%%%%%%%%%%%%%%%%%%%%
\section{ Theoretical framework}\label{sec:f-work}

Based on the pQCD factorization approach \cite{li2003},
the decay amplitude ${\cal A}(B \to M_1 M_2)$
can be written conceptually as the convolution,
\beq
{\cal A}(B \to M_1 M_2)\sim \int\!\! d^4k_1 d^4k_2 d^4k_3\ \mathrm{Tr}
\left [ C(t) \Phi_B(k_1) \Phi_{M_1}(k_2) \Phi_{M_2}(k_3)
H(k_1,k_2,k_3, t) \right ], \label{eq:con1}
\eeq
where $k_i$'s are momenta of light quarks included in each meson, and $\mathrm{Tr}$
denotes the trace over Dirac and color indices. $C(t)$ is the Wilson
coefficient evaluated at scale $t$. The function
$H(k_1,k_2,k_3,t)$ describes the four quark operator and the spectator quark connected by
a hard gluon and could be calculated perturbatively.
The function $\Phi_B$ and $\Phi_{M_i}$ are the wave functions of the initial heavy
B meson and the final light meson $M_i$,
which describe the hadronization of the quark and anti-quark into the mesons.
While the hard kernel $H$ depends on the processes considered, the wave functions
$\Phi_B$ and $\Phi_{M_i}$ are independent of the specific processes.

In the $B$ meson rest-frame, it is convenient to use light-cone coordinate $(p^+,
p^-, {\bf p}_{\rm T})$ to describe the meson's momenta:
$p^\pm = \frac{1}{\sqrt{2}} (p^0 \pm p^3)$ and ${\bf p}_{\rm T} = (p^1, p^2)$ .
Using these coordinates the $B$ meson and the two
final state meson momenta can be written as
\beq
P_B = \frac{M_B}{\sqrt{2}} (1,1,{\bf 0}_{\rm T}), \quad
P_V = \frac{M_B}{\sqrt{2}}(1,r^2_V,{\bf 0}_{\rm T}), \quad
P_P = \frac{M_B}{\sqrt{2}} (0,1-r^2_V,{\bf 0}_{\rm T}),
\eeq
respectively, here $r_V=m_V/M_B$ with $V=\rho, \omega$ or $\phi$.
 The light meson ($P=\etap$) mass has been
neglected. For the $B \to V P$ decays considered here, only the
vector meson's longitudinal part contributes to the decays, and its
polarization vector is $\epsilon_L=\frac{M_B}{\sqrt{2 M_V}}(1,-r^2_V ,0_{\rm T})$.
Putting the anti-quark momenta in $B$, $V$ and $P$ mesons as $k_1$, $k_2$, and $k_3$,
respectively,
we can choose
\beq
k_1 = (x_1 P_1^+,0,{\bf k}_{\rm 1T}), \quad
k_2 = (x_2 P_2^+,0,{\bf k}_{\rm 2T}), \quad
k_3 = (0, x_3 P_3^-,{\bf k}_{\rm 3T}).
\eeq
Then, the integration over $k_1^-$, $k_2^-$, and $k_3^+$ in
eq.(\ref{eq:con1}) will lead to
\beq
{\cal A}(B \to P V ) &\sim
&\int\!\! d x_1 d x_2 d x_3 b_1 d b_1 b_2 d b_2 b_3 d b_3 \non &&
\cdot \mathrm{Tr} \left [ C(t) \Phi_B(x_1,b_1) \Phi_V(x_2,b_2)
\Phi_P(x_3, b_3) H(x_i, b_i, t) S_t(x_i)\, e^{-S(t)} \right ],
\quad \label{eq:a2}
\eeq
where $b_i$ is the conjugate space
coordinate of $k_{iT}$. The large logarithms ($\ln m_W/t$) coming
from QCD radiative corrections to four quark operators are included
in the Wilson coefficients $C(t)$. The large double logarithms
($\ln^2 x_i$) on the longitudinal direction are summed by the
threshold resummation \cite{li02}, and they lead to $S_t(x_i)$ which
smears the end-point singularities on $x_i$. The last term,
$e^{-S(t)}$, is the Sudakov form factor which suppresses the soft
dynamics effectively \cite{li2003}.

For the studied $B \to V \etap$ decays, the weak effective Hamiltonian $H_{eff}$
can be written as \cite{buras96}
\beq
\label{eq:heff}
{\cal H}_{eff} = \frac{G_{F}}
{\sqrt{2}} \, \sum_{q=u,c}V_{qb} V_{qd}^*\left[  \left (C_1(\mu)
O_1^q(\mu) + C_2(\mu) O_2^q(\mu) \right)
+ \sum_{i=3}^{10} C_{i}(\mu) \;O_i(\mu) \right] \; .
\eeq
where $G_{F}=1.166 39\times 10^{-5} GeV^{-2}$ is the Fermi constant,
and $V_{ij}$ is the CKM matrix element, $C_i(\mu)$ are the Wilson coefficients evaluated
at the renormalization scale $\mu$ and $O_i(\mu)$ are the four-fermion operators.
For the case of
$b \to s $ transition, simply make a replacement of $d$ by $s$ in Eq.~(\ref{eq:heff})
and in the
expressions of $O_i(\mu)$ operators, which can be found easily for example in
Refs.\cite{guo07,buras96}.

In PQCD approach, the energy scale $``t"$ is chosen as the largest energy scale in
the hard kernel $H(x_i,b_i,t)$ of a given Feynman diagram, in order to
suppress the higher order corrections and improve the reliability of the perturbative
calculation.
Here, the scale $``t"$ may be larger or smaller than the $m_b$ scale.
In the range of $ t < m_b $ or $t \geq m_b$, the number of active quarks is $N_f=4$ or
$N_f=5$, respectively.
For the Wilson coefficients $C_i(\mu)$ and their renormalization group (RG) running,
they are known at NLO level currently \cite{buras96}.
The explicit expressions of the LO and NLO $C_i(\mw)$ can be found easily, for example, in
Refs.~\cite{buras96,luy01}.

When the pQCD approach at leading-order are employed, the leading order Wilson
coefficients $C_i(m_W)$, the leading order RG evolution matrix $U(t,m)^{(0)}$ from
the high scale $m$ down to $t < m$ ( for details see Eq.~(3.94) in Ref.~\cite{buras96}),
and the leading order $\alpha_s(t)$ are used:
\beq
\alpha_s(t)=\frac{4\pi}{ \beta_0 \ln \left [ t^2/ \Lambda_{QCD}^2\right]},
\eeq
where $\beta_0 = (33- 2 N_f)/3$, $\Lambda_{QCD}^{(5)}=0.225 GeV$ and
$\Lambda_{QCD}^{(4)}=0.287$ GeV.

When the NLO contributions are taken into account, however,
the NLO Wilson coefficients $C_i(m_W)$, the NLO RG evolution matrix $U(t,m,\alpha)$
(for details see Eq.~(7.22) in Ref.~\cite{buras96}) and the $\alpha_s(t)$ at two-loop level
are used:
\beq
\alpha_s(t)=\frac{4\pi}{ \beta_0 \ln \left [ t^2/ \Lambda_{QCD}^2\right]}
\cdot \left \{ 1- \frac{\beta_1}{\beta_0^2 } \cdot
\frac{ \ln\left [ \ln\left [ t^2/\Lambda_{QCD}^2  \right]\right]}{
\ln\left [ t^2/\Lambda_{QCD}^2\right]} \right \},
\label{eq:asnlo}
\eeq
where $\beta_0 = (33- 2 N_f)/3$, $\beta_1 = (306-38 N_f)/3$, $\Lambda_{QCD}^{(5)}=0.225$ GeV and
$\Lambda_{QCD}^{(4)}=0.326$ GeV.

\section{Decay amplitudes at leading order}\label{ssec:lo1}

\begin{figure}[tb]
\vspace{-3cm}
\centerline{\epsfxsize=18cm \epsffile{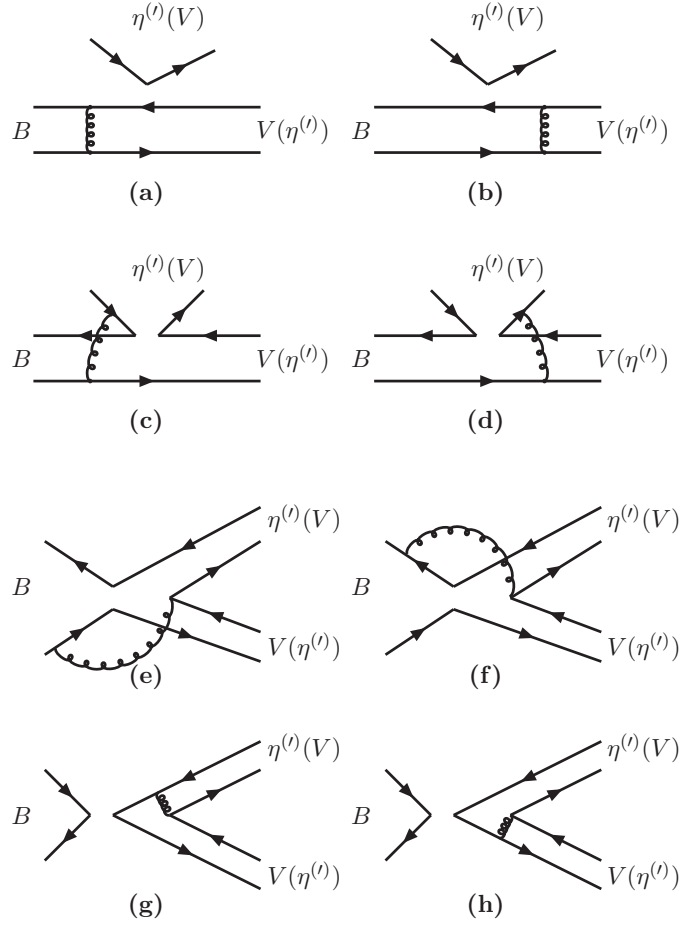}}
\vspace{-10cm}
\caption{ Feynman diagrams which may contribute to the $B\to \rho (\omega, \phi) \eta^{(\prime)}$
 decays at leading order.}
 \label{fig:fig1}
\end{figure}

In  the pQCD approach, the Feynman diagrams  as shown in Fig.~\ref{fig:fig1} may contribute
to $B \to \rho(\omega,\phi) \etap$ decays at leading order.
These decays have been studied previously in
Refs.~\cite{liu06,guodq07} by using the pQCD approach.
In this paper, we focus on
the calculations of some NLO contributions to these decays in the pQCD approach.
We firstly recalculated and confirmed the previous calculation. For the sake of
completeness, we present the relevant LO decay amplitudes in this section.

At the leading order, the total decay amplitudes
for $B \to \rho \eta$, $B^0 \to \omega \eta$, and $B^0 \to
\phi \eta$ can be written as \cite{liu06,guodq07}
\beq
{\cal M}(\rho^+ \eta) &=& F_{e\rho}\left\{\left[\xi_u
a_2-\xi_t\left(2a_3+a_4-2a_5-\frac{1}{2}a_7+\frac{1}{2}a_9-\frac{1}{2}a_{10}\right)\right]\feq
\right.\non
&&\left.-\xi_t\left(a_3-\frac{1}{2}a_9-a_5+\frac{1}{2}a_7\right)\fes\right\}-F^{P2}_{e\rho}\xi_t\left(a_6-\frac{1}{2}a_8\right)\feq
\non&&+M_{e\rho}\left\{\left[\xi_u
C_2-\xi_t\left(C_3+2C_4+2C_6+\frac{1}{2}C_8-\frac{1}{2}C_9+\frac{1}{2}C_{10}\right)\right]
\; F_1(\phi)
\right.\non
&&\left. -\xi_t
\left(C_4+C_6-\frac{1}{2}C_8-\frac{1}{2}C_{10}\right)F_2(\phi)\right\}
\non
&&
+\left(M_{a\rho}+M_e +M_a\right)\left[\xi_uC_1-\xi_t\left(C_3+C_9\right)\right]F_1(\phi)
\non
&&
-\left(M^{P1}_{a\rho}+M^{P1}_a+M^{P1}_e\right)\left(C_5+C_7\right) F_1(\phi)
\non
&&
+F_e f_{\rho}\left[\xi_u a_1-\xi_t\left(a_4+a_{10}\right)\right] F_1(\phi),
\label{eq:rhoz}
\eeq
\beq
\sqrt{2}{\cal M}( \rho^0 \eta)
&=& -F_{e\rho}\left\{\left[\xi_u
a_2-\xi_t\left(2a_3+a_4-2a_5-\frac{1}{2}a_7+\frac{1}{2}a_9-\frac{1}{2}a_{10}\right)\right]\feq
\right.\non
&&\left.-\xi_t\left(a_3-\frac{1}{2}a_9-a_5+\frac{1}{2}a_7\right)\fes\right\}-F^{P2}_{e\rho}\xi_t\left(a_6-\frac{1}{2}a_8\right)\feq
\non&&-M_{e\rho}\left\{\left[\xi_u
C_2-\xi_t\left(C_3+2C_4+2C_6+\frac{1}{2}C_8-\frac{1}{2}C_9+\frac{1}{2}C_{10}\right)\right]
F_1(\phi)
\right.\non
&&\left. -\xi_t
\left(C_4+C_6-\frac{1}{2}C_8-\frac{1}{2}C_{10}\right)F_2(\phi)\right\}
\non
&&
+\left(M_{a\rho}+M_a\right) \left[
\xi_uC_2-\xi_t\left(-C_3+\frac{3}{2}C_8+\frac{1}{2}C_9+\frac{3}{2}C_{10}\right)\right]F_1(\phi)
\non
&& -\left(M^{P1}_{a\rho} +M^{P1}_a+M^{P1}_e\right)
\;\xi_t \;\left(C_5-\frac{1}{2}C_7\right)F_1(\phi)
\non
&&
+ M_e\left[\xi_u C_2-\xi_t\left(-C_3-\frac{3}{2}C_8+\frac{1}{2}C_9
+\frac{3}{2}C_{10}\right)\right]F_1(\phi)
\non
&&
+F_e f_{\rho}\left[\xi_u
a_1-\xi_t\left(-a_4+\frac{3}{2}a_7 +
\frac{3}{2}a_9+\frac{1}{2}a_{10}\right)\right]\; F_1(\phi),
\label{eq:rho0}
\eeq
\beq
\sqrt{2}{\cal M}(\omega\eta) &=& F_{e\omega}\left\{\left[\xi_u
a_2-\xi_t\left(2a_3+a_4-2a_5-\frac{1}{2}a_7+\frac{1}{2}a_9-\frac{1}{2}a_{10}\right)\right]\feq
\right.\non
&&\left.-\xi_t\left(a_3-\frac{1}{2}a_9-a_5+\frac{1}{2}a_7\right)\fes\right\}-F^{P2}_{e\omega}\xi_t\left(a_6-\frac{1}{2}a_8\right)\feq
\non
&&-M_{e\omega}\left\{\left[\xi_u
C_2-\xi_t\left(C_3+2C_4+2C_6+\frac{1}{2}C_8-\frac{1}{2}C_9+\frac{1}{2}C_{10}\right)\right]
\; F_1(\phi)
\right.\non &&\left. -\xi_t
\left(C_4+C_6-\frac{1}{2}C_8-\frac{1}{2}C_{10}\right)F_2(\phi)\right\}
\non
&&
+\left(M_{a\omega}+M_a\right) \left[
\xi_uC_2-\xi_t\left(C_3+2C_4-\frac{1}{2}C_9+\frac{1}{2}C_{10}\right)\right]F_1(\phi)
\non
&&
-\left(M^{P1}_{a\omega}+M^{P1}_a + M^{P1}_e\right)
\; \xi_t \; \left(C_5-\frac{1}{2}C_7\right)F_1(\phi)
\non
&&
+M_e\left[\xi_u
C_2-\xi_t\left(C_3+2C_4-2C_6 -\frac{1}{2}C_8-\frac{1}{2}C_9+\frac{1}{2}C_{10}\right)\right]F_1(\phi)
\non
&&
+F_ef_{\omega}\left[\xi_u a_2-\xi_t\left(2a_3+a_4+2a_5 +\frac{1}{2}a_7+
\frac{1}{2}a_9-\frac{1}{2}a_{10}\right)\right]F_1(\phi)
\non
&&
-\left(M_a^{P2}+M_{a\omega}^{P2}\right)\; \xi_t\;
\left(2C_6+\frac{1}{2}C_8\right) F_1(\phi),
\label{eq:omega}
\eeq
\beq
 {\cal M}(\phi\eta) &=&
-F_{e}f_{\phi}\xi_t\left(a_3+a_5-\frac{1}{2}a_7-\frac{1}{2}a_9\right)F_1(\phi)
\non
&&
-M_e\xi_t\left(C_4-C_6+\frac{1}{2}C_8-\frac{1}{2}C_{10}\right)F_1(\phi)
\non
&&
-\left(M_a+M_{a\phi}\right)\xi_t\left(C_4-\frac{1}{2}C_{10}\right)F_2(\phi)\non
&&-\left(M_a^{P2}+M_{a\phi}^{P2}\right)\xi_t
\left(C_6-\frac{1}{2}C_8\right)F_2(\phi),
\label{eq:phi}
\eeq
where $\xi_u = V_{ub}^*V_{ud}$, $\xi_t = V_{tb}^*V_{td}$, and
$F_1(\phi), F_2(\phi)$ are the mixing factors of $\eta-\etar$ system
 as define in Eq.~(\ref{eq:e-ep}):
\beq
F_1(\phi)=\frac{1}{\sqrt{2}}\cos\phi , \quad F_2(\phi)=-\sin\phi .
\label{eq:m1}
\eeq
The Wilson coefficients $a_i$ appeared in the expressions of the total decay amplitude are the
combinations of the ordinary Wilson coefficients $C_i(\mu)$,
\beq
a_1(\mu)&=& C_2(\mu)+\frac{1}{3} C_1(\mu),\quad
a_2(\mu)= C_1(\mu)+\frac{1}{3} C_2(\mu),\non
a_i(\mu)&=& C_i(\mu) + \frac{C_{i\pm 1}(\mu)}{3}, for \ \ i=3,5,7,9; \ \ or \ \ 4,6,8,10.
\eeq

The individual decay amplitudes $F_{e\rho}, \cdots$ , as given  in
Eqs.~(\ref{eq:rhoz}-\ref{eq:phi}),
are obtained by evaluating individual Feynman diagrams for a
given decay mode and can be written as
\beq
F_{eV}&=& 4\sqrt{2}\pi G_F C_F m_B^4\int_0^1 d x_{1} dx_{2}\,
\int_{0}^{\infty} b_1 db_1 b_2 db_2\, \phi_B(x_1,b_1) \non & &
\times \left\{ \left[(1+x_2) \pva({\xeba}) -(1-2x_2) \rv
(\pvp(\xeba) -\pvt(\xeba))\right]
E_e(t_a)h_e(x_1,x_2,b_1,b_2)\right.\non && \left. -2\rv \pvp (\xeba)
E_e(t_a^{\prime})h_e(x_2,x_1,b_2,b_1) \right\}, \label{eq:ab}\\
F_{eV}^{P2}&=& 8\sqrt{2} G_F\pi C_F m_B^4 \int_{0}^{1}d x_{1}d
x_{2}\,\int_{0}^{\infty} b_1d b_1 b_2d b_2\, \pb(x_1) \non & &
\times
 \left\{-\re \left[ \pva(\xeba)- \rv((2+x_2) \pvp (\xeba)+x_2\pvt(\xeba))\right]
E_e(t_a)h_e (x_1,x_2,b_1,b_2)\right. \non & &\left. \
   +2\rv\re\pvp (\xeba)
 E_e(t_a^{\prime})h_e(x_2,x_1,b_2,b_1) \right\} ,\\
 M_{eV}&=& M^{P2}_{eV}= \frac{16}{\sqrt{3}}G_F\pi C_F m_B^4
\int_{0}^{1}d x_{1}d x_{2}\,d x_{3}\,\int_{0}^{\infty} b_1d b_1 b_3d
b_3\, \pb(x_1,b_1) \pea(\xsba) \non
 & &\times \left\{\left[-x_2\pva(\xeba)+2x_2\rv\pvt(\xeba)\right]
 E_e^{\prime}(t_b) h_n(x_1,x_2,x_3,b_1,b_3)\right\},
\eeq
\beq
M_{aV}&=&  \frac{16}{\sqrt{3}}G_F\pi C_F m_B^4\int_{0}^{1}d x_{1}d x_{2}\,d
x_{3}\,\int_{0}^{\infty} b_1d b_1 b_3d b_3\, \pb(x_1,b_1)\non &&
\times \left\{\left[(1- x_2)\pva(\xeba) \pea(\xsba) +\rv \re (1-x_2)
\left(\pvp(\xeba)+\pvt(\xeba) \right) \left(\pep(\xsba)-\pet(\xsba)
\right)\right.\right.\non && \left.\left.
 + \rv\re x_3\left(\pvp(\xeba)-\pvt(\xeba)
\right)\left(\pep(\xsba)+\pet(\xsba)\right)
 \right]E_a^{\prime}(t_c)h_{na}(x_1,x_2,x_3,b_1,b_3)\right.\non &&
\left.   -\left[ x_3 \pva(\xeba) \pea(\xsba) +4\rv \re
\pvp(\xeba)\pep(\xsba)-
 \rv\re(1-x_3)\left(\pvp(\xeba)+\pvt(\xeba)\right)\right.\right.\non &&
 \cdot\left.\left.\left(\pep(\xsba)-
 \pet(\xsba)\right)- \rv\re x_2
 \left(\pvp(\xeba)-\pvt(\xeba)\right)\left(\pep(\xsba)+\pet(\xsba)\right)\right]
\right.\non &&\times\left.
E_a^{\prime}(t_c^{\prime})h_{na}^{\prime}(x_1,x_2,x_3,b_1,b_3)
 \right \}\; ,
\eeq
%%%---------------------
\beq
M_{aV}^{P1}&=&  \frac{16}{\sqrt{3}}G_F\pi C_F
m_B^4\int_{0}^{1}d x_{1}d x_{2}\,d x_{3}\,\int_{0}^{\infty} b_1d b_1
b_3d b_3\, \pb(x_1,b_1)
 \non
& &\times \left\{ \left[ (1-x_2) \rv
\pea(\xsba)\left(\pvp(\xeba)+\pvt(\xeba) \right)-\re x_3
\pva(\xeba)\left( \pep(\xsba) -\pet(\xsba)\right)\right]\right.\non
&&\left.\times
E_a^{\prime}(t_c)h_{na}(x_1,x_2,x_3,b_1,b_3)-\left[-(x_2+1)\rv
\pea(\xsba)\left(\pvp(\xeba)+ \pvt(\xeba) \right) \right.\right.\non
&&\left.\left.-\re
(x_3-2)\pva(\xeba)\left(\pep(\xsba)-\pet(\xsba)\right)\right]
E_a^{\prime}(t_c^{\prime})h_{na}^{\prime}(x_1,x_2,x_3,b_1,b_3)
\right \} ,
\eeq
\beq
M_{aV}^{P2} &=&\frac{16}{\sqrt{3}}G_F\pi
C_F m_B^4\int_{0}^{1}d x_{1}d x_{2}\,d x_{3}\,\int_{0}^{\infty} b_1d
b_1 b_3d b_3\, \pb(x_1,b_1) \left\{\left[(x_2-1)\right.\right.\non
&&\left.\left.\times\pva(\xeba)\pea(\xsba)-4\re\rv\pvp(\xeba)\pep(\xsba)+\re\rv
x_2\left(\pvp(\xeba)+\pvt(\xeba)\right) \right.\right.\non
&&\left.\left. \cdot\left(\pep(\xsba)-\pet(\xsba)\right) +\rv\re
(1-x_3)\left(\pvp(\xeba)-\pvt(\xeba)\right)\left(\pep(\xsba)+\pet(\xsba)\right)
\right] \right.\non &&\left. \cdot
E_a^{\prime}(t_e)h_{na}(x_1,x_2,x_3,b_1,b_3) +
\left[x_3\pva(\xeba)\pea(\xsba)+x_3\rv\re\left(\pvp(\xeba)+\pvt(\xeba)\right)\right.\right.\quad\non
&&\left.\left.
\cdot\left(\pep(\xsba)-\pet(\xsba)\right)+\rv\re(1-x_2)\left(\pvp(\xeba)-\pvt(\xeba)\right)
\left(\pep(\xsba)+\pet(\xsba)\right) \right]\right.\non && \left.
\times
E_a^{\prime}(t_e^{\prime})h_{na}^{\prime}(x_1,x_2,x_3,b_1,b_3)\right\},
\eeq
\beq
F_{e}&=& 4\sqrt{2}G_F\pi C_F m_B^4\int_0^1 d
x_{1} dx_{2}\, \int_{0}^{\infty} b_1 db_1 b_2 db_2\, \phi_B(x_1,b_1)
\non & & \times \left\{ \left[(1+x_2) \pea({\xeba}) +(1-2x_2) \re
(\pep(\xeba) -\pet(\xeba))\right]
E_e(t_a)h_e(x_1,x_2,b_1,b_2)\right.\non && \left. +2\re \pep (\xeba)
E_e(t_a^{\prime})h_e(x_2,x_1,b_2,b_1) \right\}, \label{eq:ab1}\\
M_{e}&=& \frac{16}{\sqrt{3}}G_F\pi C_F m_B^4
\int_{0}^{1}d x_{1}d x_{2}\,d x_{3}\,\int_{0}^{\infty} b_1d b_1 b_3d
b_3\, \pb(x_1,b_1) \phi_V(\xsba) \non
 & &\times \left\{ -x_2\pea(\xeba)- 2 x_2\re \pet(\xeba)\right \}
\cdot E_e^{\prime}(t_b) h_n(x_1,x_2,x_3,b_1,b_3),\\
M_{e}^{P1}&=& -\frac{128}{\sqrt{6}}\pi C_F m_B^2\rv \int_{0}^{1}d x_{1}d x_{2}\,d
x_{3}\,\int_{0}^{\infty} b_1d b_1 b_3d b_3\, \pb(x_1,b_1)
\non
&& \times\left\{ (1-x_3)\pea(\xeba) \cdot\left(\pvp(\xsba)-\pvt(\xsba)\right)\right.\non
&& \left.
+
\re(1-x_3)\left(\pep(\xeba)+\pet(\xeba)\right)
\left(\pvp(\xsba)-\pvt(\xsba)\right)\right.
\non
&&
\left.+\re x_2\left(\pep(\xeba)-\pet(\xeba)\right)
 \left(\pvp(\xsba)+\pvt(\xsba)\right ) \right \}
\cdot E_e^{\prime}(t_b)h_n(x_1,x_2,\bar{x}_3,b_1,b_3),
\eeq
\beq
M_{a}&=&
\frac{16}{\sqrt{3}}\sqrt{2}G_F\pi C_F m_B^4\int_{0}^{1}d x_{1}d
x_{2}\,d x_{3}\,\int_{0}^{\infty} b_1d b_1 b_3d b_3\,
\pb(x_1,b_1)\non && \times \left\{\left[(1- x_2)\pea(\xeba)
\pva(\xsba) -\re \rv (1-x_2) \left(\pep(\xeba)+\pet(\xeba) \right)
\left(\pvp(\xsba)-\pvt(\xsba) \right)\right.\right.\non &&
\left.\left.
 -\re\rv x_3\left(\pep(\xeba)-\pet(\xeba)
\right)\left(\pvp(\xsba)+\pvt(\xsba)\right)
 \right]E_a^{\prime}(t_c)h_{na}(x_1,x_2,x_3,b_1,b_3)\right.\non &&
\left.   -\left[ x_3 \pea(\xeba) \pva(\xsba) -4\re \rv
\pep(\xeba)\pvp(\xsba)+
 \re\rv(1-x_3)\left(\pep(\xeba)+\pet(\xeba)\right)\right.\right.\non &&
 \cdot\left.\left.\left(\pvp(\xsba)-
 \pvt(\xsba)\right)+\re\rv x_2
 \left(\pep(\xeba)-\pet(\xeba)\right)\left(\pvp(\xsba)+\pvt(\xsba)\right)\right]
\right.\non &&\times\left.
E_a^{\prime}(t_c^{\prime})h_{na}^{\prime}(x_1,x_2,x_3,b_1,b_3)
 \right \}\; ,
 \eeq
%%%%%%%%%%%%%%%%%%%%%%%%%%%%
\beq M_{a}^{P1}&=& \frac{16}{\sqrt{3}}G_F\pi C_F m_B^4\int_{0}^{1}d
x_{1}d x_{2}\,d x_{3}\,\int_{0}^{\infty} b_1d b_1 b_3d b_3\,
\pb(x_1,b_1)
 \non
& &\times \left\{ \left[ -(1-x_2) \re
\pva(\xsba)\left(\pep(\xeba)+\pet(\xeba) \right)-\rv x_3
\pea(\xeba)\left( \pvp(\xsba) -\pvt(\xsba)\right)\right]\right.\non
&&\left.\times
E_a^{\prime}(t_c)h_{na}(x_1,x_2,x_3,b_1,b_3)-\left[(x_2+1)\re
\pva(\xsba)\left(\pep(\xeba)+ \pet(\xeba) \right) \right.\right.\non
&&\left.\left.-\rv
(x_3-2)\pea(\xeba)\left(\pvp(\xsba)-\pvt(\xsba)\right)\right]
E_a^{\prime}(t_c^{\prime})h_{na}^{\prime}(x_1,x_2,x_3,b_1,b_3)
\right \} \; ,
\eeq
%%---------------------------
\beq
M_{a}^{P2} &=&\frac{16}{\sqrt{3}}G_F\pi C_F
m_B^4\int_{0}^{1}d x_{1}d x_{2}\,d x_{3}\,\int_{0}^{\infty} b_1d b_1
b_3d b_3\, \pb(x_1,b_1) \left\{\left[(x_2-1)\right.\right.\non
&&\left.\left.\times\pea(\xeba)\pva(\xsba)+4\rv\re\pep(\xeba)\pvp(\xsba)-\rv\re
x_2\left(\pep(\xeba)+\pet(\xeba)\right) \right.\right.\non
&&\left.\left. \cdot\left(\pvp(\xsba)-\pvt(\xsba)\right) -\re\rv
(1-x_3)\left(\pep(\xeba)-\pet(\xeba)\right)\left(\pvp(\xsba)+\pvt(\xsba)\right)
\right] \right.\non &&\left. \cdot
E_a^{\prime}(t_e)h_{na}(x_1,x_2,x_3,b_1,b_3) +
\left[x_3\pea(\xeba)\pva(\xsba)-x_3\re\rv\left(\pep(\xeba)+\pet(\xeba)\right)\right.\right.\quad\non
&&\left.\left.
\cdot\left(\pvp(\xsba)-\pvt(\xsba)\right)-\re\rv(1-x_2)\left(\pep(\xeba)-\pet(\xeba)\right)
\left(\pvp(\xsba)+\pvt(\xsba)\right) \right]\right.\non && \left.
\times
E_a^{\prime}(t_e^{\prime})h_{na}^{\prime}(x_1,x_2,x_3,b_1,b_3)\right\},
 \eeq
Here $\rv=m_V/m_B$ is the mass ratio with $(m_V=m_\rho, m_\omega,m_\phi)$;
$C_F=4/3$ is a color factor. The evolution functions $E(t_i)$ and hard function $h_j(x_i,b_i)$
are displayed in Appendix \ref{sec:aa}.

The decay amplitudes for $B \to \rho \eta^{\prime}$, $B^0 \to \omega
\eta^{\prime}$, and $B^0 \to \phi \eta^{\prime}$ decays can be
obtained easily from Eqs.(\ref{eq:rhoz}) to (\ref{eq:phi}) by the
following replacements
\beq
f_\eta^{q}, f_\eta^s &\longrightarrow& f_{\eta^\prime}^q, f_{\eta^\prime}^s, \\
F_1(\phi) &\longrightarrow & F'_1(\phi) =\frac{1}{\sqrt{2}}\sin\phi,
\\ F_2(\phi) &\longrightarrow & F'_2(\phi) =\cos\phi.
\eeq

%%%%%%%%%%%%%%%%%%%%%%%%%%%%%%%%%%%%%%%%%%%%%%%%%%%%%%%%%%%

\section{Next-to-leading contributions}\label{sec:nlo}

The power counting in the pQCD approach \cite{nlo05} is different
from that in the QCDF approach\cite{bbns99}.
Here the term NLO means that the decay amplitude  is proportional to $\alpha^2_s(\mu)$.
We here indeed consider the partial NLO contributions only: those
from the vertex corrections, the
quark-loops and chromo-magnetic penguins. The NLO contributions from hard-spectator
and annihilation diagrams are not known at present.
When compared with the previous LO calculations
in pQCD \cite{guo07}, the following NLO contributions will be included:

\begin{enumerate}

\item
The LO Wilson coefficients $C_i(\mw)$ will be replaced by those at NLO level in NDR scheme
\cite{buras96}. As mentioned in last section,  the strong coupling constant
$\alpha_s(t)$ at two-loop level as given in Eq.~(\ref{eq:asnlo}), and the NLO RG evolution
matrix $U(t,m,\alpha)$, as defined in Ref.~\cite{buras96}, will be used here:
\beq
U(m_1,m_2,\alpha) = U(m_1,m_2) + \frac{\alpha}{4\pi} R(m_1,m_2)
\label{eq:um1m2a}
\eeq
where the function  $U(m_1,m_2)$ and $R(m_1,m_2)$ represent the QCD and QED evolution and
have been defined in Eq.~(6.24) and (7.22) in Ref.~\cite{buras96}. We also introduce a cut-off
 $\mu_0 = 1$ GeV for the hard  scale ``t" in the final integration.

\end{enumerate}

\subsection{Vertex corrections}

\begin{figure}[tb]
\vspace{-3cm}
\centerline{\epsfxsize=19 cm \epsffile{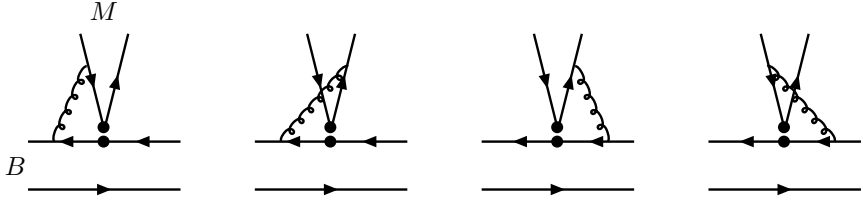}}
\vspace{-21cm}
\caption{NLO vertex corrections to the factorizable amplitudes. }
\label{fig:fig2}
\end{figure}

The vertex corrections to the factorizable emission diagrams, as illustrated by
Fig.~\ref{fig:fig2}, have been calculated years ago in the QCD factorization
appeoach\cite{bbns99,npb675}.
According to Ref.~\cite{nlo05}, the difference of the calculations induced by
considering or not considering the parton transverse momentum is rather small, say
less than $10\%$, and therefore can be neglected.
Consequently, one can use the vertex corrections as given in Ref.~\cite{npb675} directly.
The vertex corrections can be absorbed into the re-definition of the Wilson coefficients
$a_i(\mu)$ by adding a vertex-function $V_i(M)$ to them
\cite{bbns99,npb675}
\beq
a_i(\mu)&\to & a_i(\mu) +\frac{\alpha_s(\mu)}{4\pi}C_F\frac{C_i(\mu)}{3} V_i(M),\ \ for
\ \ i=1,2; \non
a_j(\mu)&\to & a_j(\mu)+\frac{\alpha_s(\mu)}{4\pi}C_F\frac{C_{j\pm
1}(\mu)}{N_c} V_j(M), \ \  for \ \ j=3-10,
\label{eq:aimu-2}
\eeq
where M is the meson emitted from the weak vertex. When $M$ is a
pseudo-scalar meson, the vertex functions $V_{i}(M)$ are given ( in the NDR
scheme) in Refs.~\cite{nlo05,npb675}:
\beq
V_i(M)&=&\left\{ \begin{array}{cc}
12\ln\frac{m_b}{\mu}-18+\frac{2\sqrt{2N_c}}{f_M}\int_{0}^{1}dx\phi_M^A(x)g(x),
& {\rm for}\quad i= 1-4,9,10,\\
-12\ln\frac{m_b}{\mu}+6-\frac{2\sqrt{2N_c}}{f_M}\int_{0}^{1}dx\phi_M^A(x)g(1-x),
&{\rm for}\quad i= 5,7,\\
-6+\frac{2\sqrt{2N_c}}{f_M}\int_{0}^{1}dx\phi_M^P(x)h(x),
&{\rm for} \quad i= 6,8,\\
\end{array}\right.
\label{eq:vim}
\eeq
where $f_M$ is the decay constant of the meson M; $\phi_M^A(x)$ and $\phi_M^P(x)$ are
the twist-2 and twist-3 distribution amplitude of the meson M, respectively.
For a vector meson V, $\phi_M^A(\phi_M^P)$ is replaced by $\phi_V (\phi_V^s)$
and $f_M$ by $f_V^T$ in the third line of the above formulas. The hard-scattering
functions $g(x)$ and $h(x)$ in Eq.~(\ref{eq:vim}) are:
\beq
g(x)&=& 3\left(\frac{1-2x}{1-x}\ln x-i\pi\right)\non
&& +\left[2Li_2(x)-\ln^2x+\frac{2\ln
x}{1-x}-\left(3+2i\pi\right)\ln x -(x\leftrightarrow 1-x)\right],\\
h(x)&=&2Li_2(x)-\ln^2x-(1+2i\pi)\ln x-(x\leftrightarrow 1-x),
\eeq
where $Li_2(x)$ is the dilogarithm function. As shown in Ref.~\cite{nlo05},
the $\mu$-dependence of the Wilson coefficients $a_i(\mu)$ will be improved generally
by the inclusion of the vertex corrections.

%%%%%%%%%%%%%%%%%%%%%%%%%%%%%%%%%%%%%%%%%%%%%%%%%%%%%%%%%%%%%%%%%%%%%%%%%%%%

\subsection{Quark loops}

\begin{figure}[tb]
\vspace{-4cm}
\centerline{\epsfxsize=19 cm \epsffile{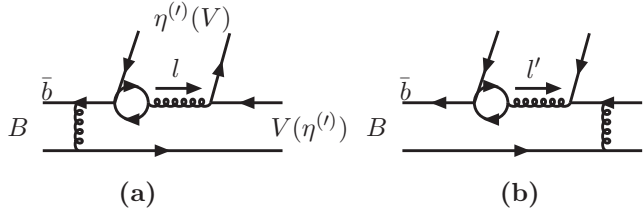}}
\vspace{-20cm}
\caption{Quark-loop diagrams contributing to $B \to \rho(\omega)\etap$ decays.}
\label{fig:fig3}
\end{figure}

The contribution from the so-called ``quark-loops" is a kind of penguin correction
with the four quark operators insertion, as illustrated by Fig.~\ref{fig:fig3}.
In fact this is generally called BSS mechanism\cite{bss79}, which
plays a very important role in producing the CP violation in the QCDF/SCET approaches.
We here include quark-loop amplitude from the
operators $O_{1,2}$ and $O_{3-6}$ only. The quark loops from $O_{7-10}$
will be neglected due to their smallness.

For the $b\to d$ transition, the contributions from the various
quark loops are described by the effective Hamiltonian $H_{eff}^{(ql)}$ \cite{nlo05},
\beq
H_{eff}^{(ql)}&=&-\sum\limits_{q=u,c,t}\sum\limits_{q{\prime}}\frac{G_F}{\sqrt{2}}
V_{qb}V_{qd}^{*}\frac{\alpha_s(\mu)}{2\pi}\; C^{(q)}(\mu,l^2)\; \left(\ov{d}\gamma_\rho
\left(1-\gamma_5\right)T^ab\right)\left(\ov{q}^{\prime}\gamma^\rho
T^a q^{\prime}\right),
\eeq
where $l^2$ being the invariant mass of the gluon, which connects the quark loops with the
$\ov{q}' q$ pair as shown in Fig.~\ref{fig:fig3}.
The functions $C^{(q)}(\mu,l^2)$ can be written as
\beq
C^{(q)}(\mu,l^2)&=&\left[G^{(q)}(\mu,l^2)-\frac{2}{3}\right]C_2(\mu),
\label{eq:qlc}
\eeq
for $q=u,c$ and
\beq
C^{(t)}(\mu,l^2)&=&\left[G^{(s)}(\mu,l^2)-\frac{2}{3}\right]
C_3(\mu)+\sum\limits_{q{\prime\prime}=u,d,s,c}G^{(q^{\prime\prime})}(\mu,l^2)
\left[C_4(\mu)+C_6(\mu)\right].\label{eq:eh}
\eeq
The integration function $G^{(q)}(\mu,l^2)$ for the loop of the
quarks $q=(u, d, s, c)$ is defined as \cite{nlo05}
\beq
G^{(q)}(\mu,l^2)&=&-4\int_{0}^{1}dx\; x(1-x)\ln \frac{m_q^2-x(1-x)l^2}{\mu^2},
\eeq
where $m_q$ is the quark mass. The explicit  expressions of the function
$G^{(q)}(\mu,l^2)$ after the integration can be found, for example, in
Ref.~\cite{nlo05}.

It is straightforward to calculate the decay amplitude for Fig.\ref{fig:fig3}a and
\ref{fig:fig3}b. For the case of $B \to V $ or $B \to \eta$ transition, we find
two kinds of topological decay amplitudes:
\beq
M^{(q)}_{V\eta}&=&-16m_B^2\frac{C^2_F}{\sqrt{2N_c}}\int_{0}^{1}d
x_{1}d x_{2}\,d x_{3}\,\int_{0}^{\infty} b_1d b_1 b_2d
b_2\,\phi_B(x_1,b_1)\left\{\left[\left(1+x_2\right)
 \pva(\xeba)\pea(\xsba)\right.\right.\non && \left.\left.
-\rv\left(1-2x_2\right)\left(\pvp(\xeba)-\pvt(\xeba)\right)\pea(\xsba)-
2\re\pva(\xeba)\pep(\xsba)+2\rv\re\left(\left(2+x_2\right)\right.\right.\right.\non
&&\left.\left.\left. \cdot\pvp(\xeba)+
x_2\pvt(\xeba)\right)\pep(\xsba)\right]E^{(q)}(t_q,l^2)h_e(x_2,x_1,b_2,b_1)
+\left[-2\rv\pvp(\xeba)\pea(\xsba)\right.\right.\non && \left.\left.
+4\rv\re\pvp(\xeba)\pep(\xsba)\right]E^{(q)}(t^{\prime}_q,l^{\prime
2}) h_e(x_1,x_2,b_1,b_2)\right\}, \label{eq:qlmp}
\eeq
for $B \to V$ transition, and
\beq
M^{(q)}_{\eta V}&=&-\frac{4}{\sqrt{3}}G_F C_F^2m_B^4\int_{0}^{1}d x_{1}d x_{2}\,d
x_{3}\,\int_{0}^{\infty} b_1d b_1 b_2d
b_2\,\phi_B(x_1,b_1)\left\{\left[\left(1+x_2\right)
 \pea(\xeba)\pva(\xsba)\right.\right.\non && \left.\left.
+\re\left(1-2x_2\right)\left(\pep(\xeba)-\pet(\xeba)\right)\pva(\xsba)-
2\rv\pea(\xeba)\pvp(\xsba)-2\re\rv\left(\left(2+x_2\right)\right.\right.\right.\non
&&\left.\left.\left. \cdot\pep(\xeba)+
x_2\pet(\xeba)\right)\pvp(\xsba)\right]E^{(q)}(t_q,l^2)h_e(x_2,x_1,b_2,b_1)
+\left[2\re\pep(\xeba)\pva(\xsba)\right.\right.\non && \left.\left.
-4\re\rv\pep(\xeba)\pvp(\xsba)\right]E^{(q)}(t^{\prime}_q,l^{\prime
2}) h_e(x_1,x_2,b_1,b_2)\right\}\label{eq:qlmp3},
\eeq
for $B \to \eta$ transition. Here V represents $\rho,\omega$, or $\phi$ meson, and
$\re =m_\eta/m_B, \rv=m_V/m_B$. The evolution factors take the form of
\beq
E^{(q)}(t,l^2)&=& C^{(q)}(t,l^2)\;\alpha_s^2(t)\cdot \exp\left [ -S_{ab} \right ],
\eeq
with the Sudakov factor $S_{ab}$ and the hard function $h_e(x_1,x_2,b_1,b_2) $
as given in Eq.~(\ref{eq:sab}) and Eq.~(\ref{eq:he3}) respectively, and finally
the hard scales and the gluon invariant masses are
\beq
t_{q}&=& {\rm max}(\sqrt{x_2}m_B,\sqrt{x_1x_2}m_B,\sqrt{(1-x_2) x_3}m_B,1/b_1,1/b_2);,\non
t_{q}^{\prime}&=& {\rm max}(\sqrt{x_1}m_B,\sqrt{x_1 x_2}m_B,\sqrt{|x_3-x_1|}m_B,1/b_1,1/b_2),
\label{eq:tq-tqp}\\
l^2          &=& (1-x_2) x_3 m_B^2 - |\bfkk_{\rm 2T} -\bfkk_{\rm 3T} |^2
\approx (1-x_2) x_3 m_B^2 , \non
l^{\prime 2} &=& (x_3-x_1) m_B^2- |\bfkk_{\rm 1T} -\bfkk_{\rm 3T} |^2
\approx (x_3-x_1) m_B^2.
\eeq
For $B \to V \etar$ decays, we find  the similar results  by making appropriate replacements,
such as $r_\eta \to r-\etar$, etc.

Finally, the total ``quark-loop" contribution to the considered $B \to V \etap$ decays
with $V = \rho, \omega$ can be written as
\beq
M_{V\etap}^{(ql)} &=&  <V\etap|{\cal H}_{eff}^{ql}|B>
= \sum_{q=u,c,t} \lambda_q \; \left [ M^{(q)}_{V\etap} + M^{(q)}_{\etap V} \right],
\eeq
where $\lambda_q = V_{qb}V_{qd}^{*}$. The quark-loops do not contribute to $B \to \phi\etap$ decays.

It is note that the quark-loop corrections are mode dependent. The
assumption of a constant gloun invariant mass in FA introduces a
large theoretical uncertainty as making predictions. In the PQCD approach,
the gluon invariant mass is related to the parton
momenta unambiguously.

\subsection{Magnetic penguins}

\begin{figure}[tb]
\vspace{-5cm}
\centerline{\epsfxsize=20 cm \epsffile{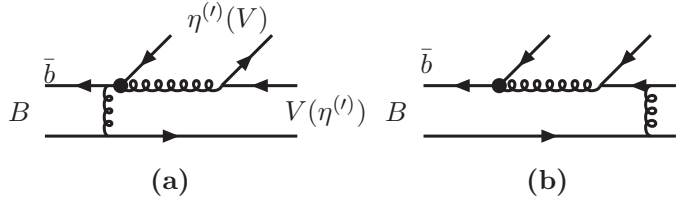}}
\vspace{-20.5cm}
\caption{Chromo-Magnetic penguin ($O_{8g}$) diagrams contributing to
$B \to \rho(\omega) \etap$ decays.}
\label{fig:fig4}
\end{figure}

As illustrated by Fig.~\ref{fig:fig4}, the chromo-magnetic penguin operator $O_{8g}$ also
contribute to $B \to V \etap$ decays at NLO level.
The corresponding weak effective Hamiltonian contains the $b\to d g$ transition,
\beq
{\cal H}_{eff}^{cmp}&=&-\frac{G_F}{\sqrt{2}}V_{tb}V_{td}^* \; C_{8g}^{eff} \; O_{8g},
\label{eq:heffo8g}
\eeq
with the chromo-magnetic penguin operator,
\beq
O_{8g}&=&\frac{g_s}{8\pi^2}m_b\; \ov{d}_i \sigma^{\mu\nu}(1+\gamma_5) T^a_{ij}G^a_{\mu\nu}
b_j,
\label{eq:o8g}
\eeq
where $i,j$ being the color indices of quarks. The corresponding effective Wilson
coefficient $C_{8g}^{eff}= C_{8g} + C_5$ \cite{nlo05}.

In Ref.~\cite{o8g2003}, the authors calculated the chromo-magnetic penguin contributions
to $B \to \phi K$ decays using the pQCD approach. They considered nine
chromo-magnetic penguin diagrams corresponding to the non-local operator
$O_{8g}^\prime$, as given in Eq.~(2.3) of Ref.~\cite{o8g2003}, generated by operator
$O_{8g}$ as defined in Eq.~(\ref{eq:o8g}).
The first two Feynman diagrams (a) and (b) in Ref.~\cite{o8g2003}
are the same as Figs.~\ref{fig:fig4}a and \ref{fig:fig4}b here.
According to Ref.~\cite{o8g2003}, the diagrams (a) and (b) dominate, while other seven
diagrams are small or negligible.
It is therefore reasonable for us to consider the NLO contributions induced by the
diagrams (a) and (b) only, for the sake of simplicity.

The decay amplitude for Figs.~\ref{fig:fig4}a and \ref{fig:fig4}b can be written  as
\beq
M^{(g)}_{V\eta}&=&16m^4_B\frac{C^2_F}{2\sqrt{N_c}}\int_{0}^{1}d
x_{1}d x_{2}\,d x_{3}\,\int_{0}^{\infty} b_1d b_1 b_2d
b_2\,\phi_B(x_1,b_1)\left\{\left[-\left(1-x_2\right)\left\{2\pva(\xeba)-
\rv\right.\right.\right. \non &
&\left.\left.\left.\left.\cdot(3\pvp(\xeba)-\pvt(\xeba)\right)-\rv
x_2
\left(\pvp(\xeba)+\pvt(\xeba)\right)\right\}\pea(\xsba)+\re\left(1+x_2\right)x_3\pva(\xeba)
\right.\right.\non
 &&\left.\left.\cdot\left(3
\pep(\xsba)+\pet(\xsba)\right)-\rv\re\left(1-x_2\right)\left(\pvp(\xeba)+\pvt(\xeba)\right)\left
(3\pep(\xsba)-\pet(\xsba)\right)\right.\right.\non
&&\left.\left.-\rv\re
x_3\left(1-2x_2\right)\left(\pvp(\xeba)-\pvt(\xeba)\right)\left(3\pep(\xsba)+\pet(\xsba)\right)\right]
\right.\non &&\left.\cdot
E_g(t_q)h_g(A,B,C,b_1,b_2,b_3,x_2)-E_g(t_q^{\prime})h_g(A^{\prime},B^{\prime},C^{\prime},b_2,b_1,b_3,x_1)\right.\non&&\left.\cdot\left[-4\rv\pvp(\xeba)\pea(\xsba)+2\rv\re
x_3\pvp(\xeba)\left(3\pep(\xsba)+\pet(\xsba)\right)\right]\right\},
\label{eq:mpp}
\eeq
for the case of $B \to V$ transition, and
\beq
M^{(g)}_{\eta
V}&=&\frac{4}{\sqrt{3}}G_F C_F^2m_B^6\int_{0}^{1}d x_{1}d x_{2}\,d
x_{3}\,\int_{0}^{\infty} b_1d b_1 b_2d
b_2\,\phi_B(x_1,b_1)\left\{\left[-\left(1-x_2\right)\left\{2\pea(\xeba)+
\re\right.\right.\right. \non &
&\left.\left.\left.\left.\cdot(3\pep(\xeba)-\pet(\xeba)\right)+\re
x_2
\left(\pep(\xeba)+\pet(\xeba)\right)\right\}\pva(\xsba)+\rv\left(1+x_2\right)x_3\pea(\xeba)
\right.\right.\non
 &&\left.\left.\cdot\left(3
\pvp(\xsba)+\pvt(\xsba)\right)+\re\rv\left(1-x_2\right)\left(\pep(\xeba)+\pet(\xeba)\right)\left
(3\pvp(\xsba)-\pvt(\xsba)\right)\right.\right.\non
&&\left.\left.+\re\rv
x_3\left(1-2x_2\right)\left(\pep(\xeba)-\pet(\xeba)\right)\left(3\pvp(\xsba)+\pvt(\xsba)\right)\right]
\right.\non &&\left.\cdot
E_g(t_q)h_g(A,B,C,b_1,b_2,b_3,x_2)-E_g(t_q^{\prime})h_g(A^{\prime},B^{\prime},C^{\prime},b_2,b_1,b_3,x_1)
\right.\non&&\left.\cdot\left[4\re\pep(\xeba)\pva(\xsba)-2\re\rv
x_3\pep(\xeba)\left(3\pvp(\xsba)+\pvt(\xsba)\right)\right]\right\}.
\label{eq:mpp2}
\eeq
for the case of $B \to \eta$ transition. Here the hard scale $t_q$ and $t_q^\prime$ are
the same as in Eq.~(\ref{eq:tq-tqp}). The evolution factor $E_g(t)$ in Eqs.~(\ref{eq:mpp})
and (\ref{eq:mpp2}) is of the form
\beq
E_g(t)&=& C_{8g}^{eff}(t)\; \alpha_s^2(t)\cdot \exp\left [ -S_{mg}\right ],
\eeq
with the Sudakov factor $S_{mg}$ and the hard  function $h_g$,
\beq
S_{mg}(t) &=& s\left(x_1 m_B/\sqrt{2}, b_1\right)
 +s\left(x_2 m_B/\sqrt{2}, b_2\right)
+s\left((1-x_2) m_B/\sqrt{2}, b_2\right) \non
 && +s\left(x_3
m_B/\sqrt{2}, b_3\right) +s\left((1-x_3) m_B/\sqrt{2}, b_3\right)
\non
 &
&-\frac{1}{\beta_1}\left[\ln\frac{\ln(t/\Lambda)}{-\ln(b_1\Lambda)}
+\ln\frac{\ln(t/\Lambda)}{-\ln(b_2\Lambda)}+\ln\frac{\ln(t/\Lambda)}{-\ln(b_3\Lambda)}\right],
\label{eq:smg}
\eeq
\beq
h_g(A,B,C,&&b_1,b_2,b_3,x_i)=-S_t(x_i) \; K_0(Bb_1) \; K_0(Cb_3)\non
&& \cdot
\int_{0}^{\pi/2}d\theta \tan\theta\;
\cdot J_0(Ab_1\tan\theta)J_0(Ab_2\tan\theta)J_0(Ab_3\tan\theta), \ \
\eeq
where the functions $K_0(x)$ and $J_0(x)$ are the
Bessel functions, the form factor
$S_t(x_i)$ with $i=1,2$ has been given in Eq.~(\ref{eq:stxi}), and
the invariant masses $A^{(\prime)}, B^{(\prime)}$ and $C^{(\prime)}$ of the virtual
quarks and gluons are of the form
\beq
A &=&\sqrt{x_2}m_B,\quad B=B^{\prime}=\sqrt{x_1 x_2}m_B,\quad C=i\sqrt{(1-x_2)x_3}m_B,\non
A^{\prime}&=&\sqrt{x_1}m_B,\quad C^{\prime}=\sqrt{|x_1-x_3|}m_B.
\eeq
For $B \to V \etar$ decays, we find the similar results by making appropriate replacements.

The total ``chromo-magnetic penguin" contribution to the considered $B \to V\etap$ decays
can therefore be written as
\beq
M_{V\etap}^{(cmp)} &=&  <V\etap |{\cal H}_{eff}^{cmp}|B>
= \lambda_t \; \left [ M^{(g)}_{V\etap} + M^{(g)}_{\etap V} \right],
\eeq
where $\lambda_t = V_{tb}V_{td}^{*}$. Again, the chromo-magnetic penguins do not contribute
to $B \to \phi \etap$ decays.

\section{Numerical results and Discussions}\label{sec:n-d}

Using the wave functions and the central values of
relevant input parameters as given in Appendix \ref{sec:wf2}, we firstly
find the numerical values of the corresponding form factors at zero momentum transfer:
\beq
A_0^{B\to \rho}(q^2=0)&=& 0.32_{-0.04}^{+0.05}(\omega_b),\non
A_0^{B\to \omega}(q^2=0)&=& 0.29_{-0.03}^{+0.04}(\omega_b),\non
F^{B\to\etap}_0(q^2=0)&=&0.22\pm{0.03}(\omega_b), \eeq
for $\omega_b=0.40\pm0.04$GeV, which agree well with those obtained in
QCD sum rule calculations.

\subsection{Branching ratios}

For a general charmless two-body decays $B \to V \etap$, the branching ratio
can be written in general as
\beq
Br(B\to V \etap) &=& \tau_B\; \frac{1}{16\pi m_B}\; \left | \calm \right|^2
\eeq
where $\tau_B$ is the lifetime of the B meson, and the decay amplitude is the form of
\beq
\calm= <V\etap| \calh_{eff} +\calh_{eff}^{(ql)} + \calh_{eff}^{(cmp)}|B>.
\eeq

Using  the wave functions and the input parameters as specified in
previous sections, it is straightforward  to calculate the CP-averaged branching
ratios for the  considered decays, which are listed in Table \ref{table1}. For
comparison, we also list the corresponding updated
experimental results \cite{pdg2006,hfag} and numerical results evaluated in
the framework of the QCD factorization (QCDF) \cite{npb675}.

\begin{table}[thb]
\begin{center}
\caption{ The pQCD predictions for the branching ratios (in unit of $10^{-6}$).
The label $\rm{LO_{NLOWC}}$ means the LO results
with the NLO Wilson coefficients, and +VC, +QL, +MP, NLO means the
inclusion  of the vertex corrections, the quark loops, the magnetic
penguin, and all the considered NLO corrections, respectively.}
\label{table1}
\vspace{0.2cm}
\begin{tabular}{l  |c |c c c c |c |c c } \hline \hline
 Mode&  LO & $LO_{{\rm NLOWC}}$ & +VC & +QL & +MP  & NLO &Data &QCDF\\
\hline
$ B^\pm \to \rho^\pm\eta$ & 6.9& 7.4& 6.8 & 7.5& 7.2 &6.7&$5.4\pm1.2$&$9.4^{+5.9}_{-4.8}$    \\
$B^\pm \to \rho^\pm\eta^{\prime}$ & 5.2& 4.8&4.6&4.9&4.7&4.6&$9.1_{-2.8}^{+3.7}$&$6.3^{+4.0}_{-3.3}$ \\
$B^0 \to \rho^0\eta$ &0.08 & 0.08&0.19&0.16&0.12&0.13& $<1.5$&$0.03^{+0.17}_{-0.10}$\\
$B^0 \to \rho^0\eta^{\prime}$ &0.05&0.04&0.13&0.06&0.04&0.10& $<1.3$&$0.01^{+0.12}_{-0.06}$ \\
$B^0 \to \omega\eta$ &0.22&0.34&0.67&0.33&0.25&0.71& $<1.9$& $0.31^{+0.46}_{-0.27}$ \\
$B^0 \to \omega\eta^{\prime}$ &0.12&0.18&0.52&0.19&0.15&0.55& $<2.2$&$0.20^{+0.34}_{-0.18}$ \\
$B^0 \to \phi\eta$ &0.001&0.002&0.011&--&--&0.011& $<0.6$&0.001 \\
$B^0 \to \phi\eta^{\prime}$ &0.096&0.053&0.017&--&--&0.017& $<0.5$&0.001 \\
 \hline \hline
\end{tabular}
\end{center}\end{table}

It is worth stressing  that the theoretical predictions in the pQCD
approach still have relatively large theoretical errors induced by the
large uncertainties of many input parameters, such as
$\omega_b$, Gegenbauer coefficient $a_2$, the CKM angle $\alpha$ and $m_s$.
The pQCD predictions with the major theoretical errors
for the branching ratios of the decays under consideration
are the following
\beq
Br(\ B^\pm \to \rho^\pm \eta) &=& \left [6.7
^{+2.2}_{-1.5}(\omega_b)^{+1.0}_{-0.9}(\mu_0) ^{+0.5}_{-0.4}(\alpha)^{+0.7}_{-0.5}(a_{2})
^{+0.1}_{-0.0}(a_{2\rho})\right ] \times 10^{-6}
,\non
Br(\ B^\pm \to \rho^\pm \eta^{\prime}) &=&\left [4.6
^{+1.4}_{-1.1}(\omega_b)^{+0.5}_{-0.7}(\mu_0) ^{+0.2}_{-0.3}(\alpha)\pm0.4(a_{2})
^{+0.0}_{-0.1}(a_{2\rho})\right ] \times 10^{-6}
,\non
Br(\ B^0 \to \rho^0 \eta) &=& \left [1.3
^{+0.4}_{-0.2}(\omega_b) ^{+1.1}_{-0.4}(\mu_0) ^{+0.1}_{-0.0}(\alpha)^{+0.1}_{-0.0}(a_{2})
^{+0.2}_{-0.1}(a_{2\rho})\right ] \times 10^{-7}
 ,\non
Br(\ B^0 \to \rho^0 \eta^{\prime}) &=& \left [1.0
^{+0.3}_{-0.2}(\omega_b)^{+0.3}_{-0.4}(\mu_0)  \pm 0.2(\alpha)^{+0.0}_{-0.1}(a_{2})
\pm0.1(a_{2\rho})\right ] \times 10^{-7},\non
Br(\ B^0 \to \omega \eta) &=& \left [7.1
^{+1.7}_{-1.3}(\omega_b) ^{+2.6}_{-1.8}(\mu_0)  ^{+0.1}_{-0.2}(\alpha)^{+1.7}_{-1.4}(a_{2})
^{+1.0}_{-0.8}(a_{2\omega})\right ] \times 10^{-7},\non
Br(\ B^0 \to \omega \eta^{\prime}) &=& \left [5.5
^{+1.3}_{-1.1}(\omega_b)^{+2.1}_{-1.6}(\mu_0)  ^{+1.2}_{-1.0}(\alpha)^{+1.3}_{-1.2}(a_{2})
^{+0.8}_{-0.7}(a_{2\omega})\right ] \times 10^{-7},\non
Br(\ B^0 \to \phi \eta) &=& \left [1.1
\pm 0.1 (\omega_b)^{+5.9}_{-0.8}(\mu_0)  ^{+0.4}_{-0.2}(m_s)^{+0.1}_{-0.2}(a_{2})
^{+0.30}_{-0.28}(a_{2\phi})\right ] \times 10^{-8},\non
Br(\ B^0 \to \phi \eta^{\prime}) &=& \left [1.7
\pm0.2(\omega_b)^{+15.3}_{-0.9}(\mu_0)  ^{+0.9}_{-0.4}(m_s)\pm0.1(a_{2})
\pm0.1(a_{2\phi})\right ] \times 10^{-8} \label{eq:etap11},
\eeq
where the major errors are induced by the uncertainties of $\omega_b=0.4
\pm 0.04$ GeV, $\mu_0=1.0\pm 0.5$ GeV, $\alpha = 100^\circ \pm 20^\circ$, $m_s=130\pm30$
MeV, Gegenbauer coefficients $a_2=0.115\pm0.115$,
$a_{2\rho}=a_{2\omega}=0.15\pm0.15$ and $a_{2\phi}=0.2\pm0.2$,
respectively.

\begin{figure}[tb]
\begin{center}
\includegraphics[scale=0.7]{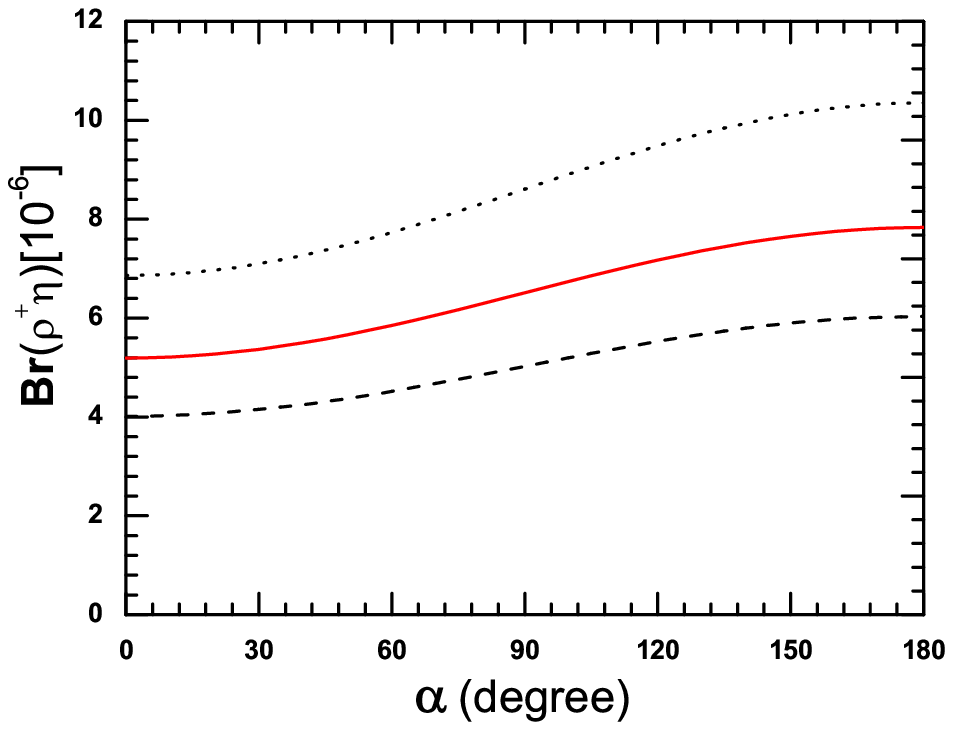}
\includegraphics[scale=0.7]{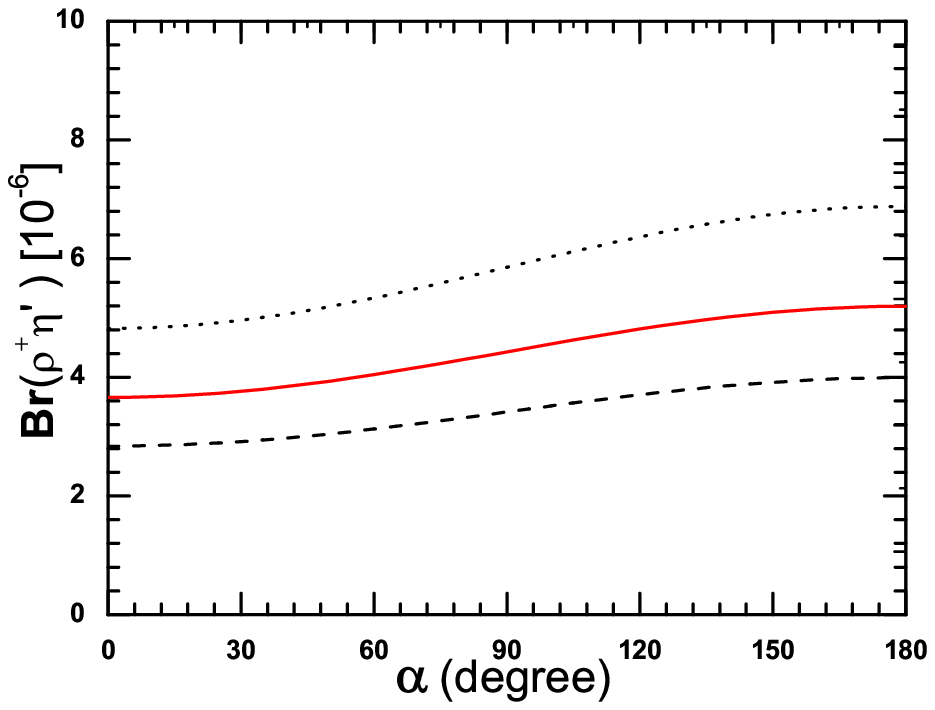}
\vspace{0.3cm}
\caption{The $\alpha$ dependence of the branching ratios of $B^+\to\rho^+\etap$ decays for
$\omega_b=0.36$ GeV (dotted curve), 0.40 GeV (solid curve) and 0.44
GeV (dashed curve).}
\label{fig:rozbralf}
\end{center}
\end{figure}

\begin{figure}[tb]
\begin{center}
\includegraphics[scale=0.7]{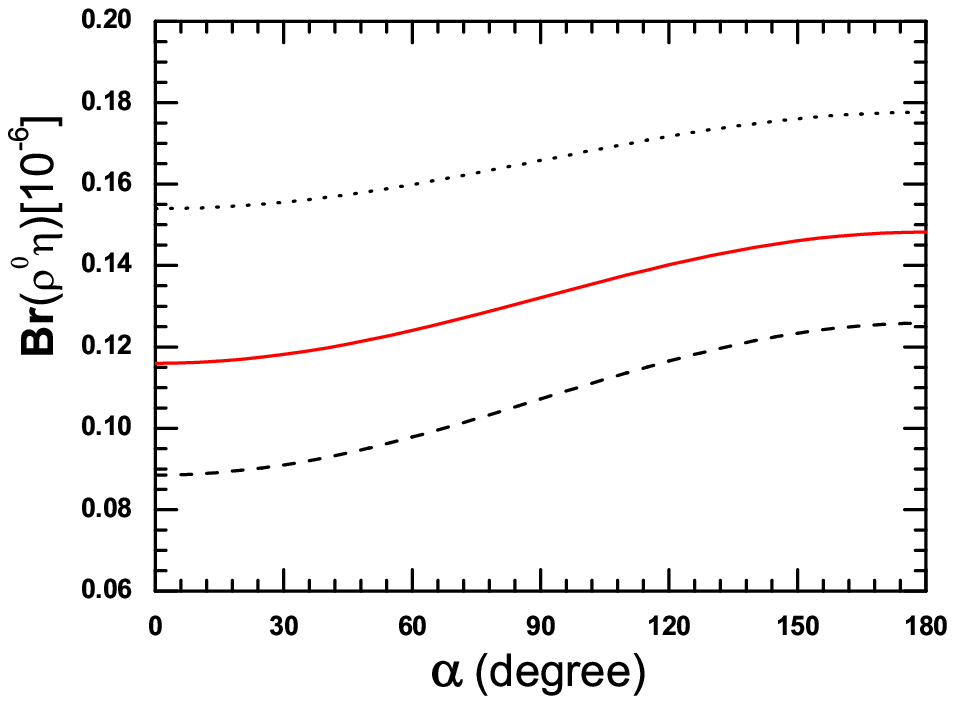}
\includegraphics[scale=0.7]{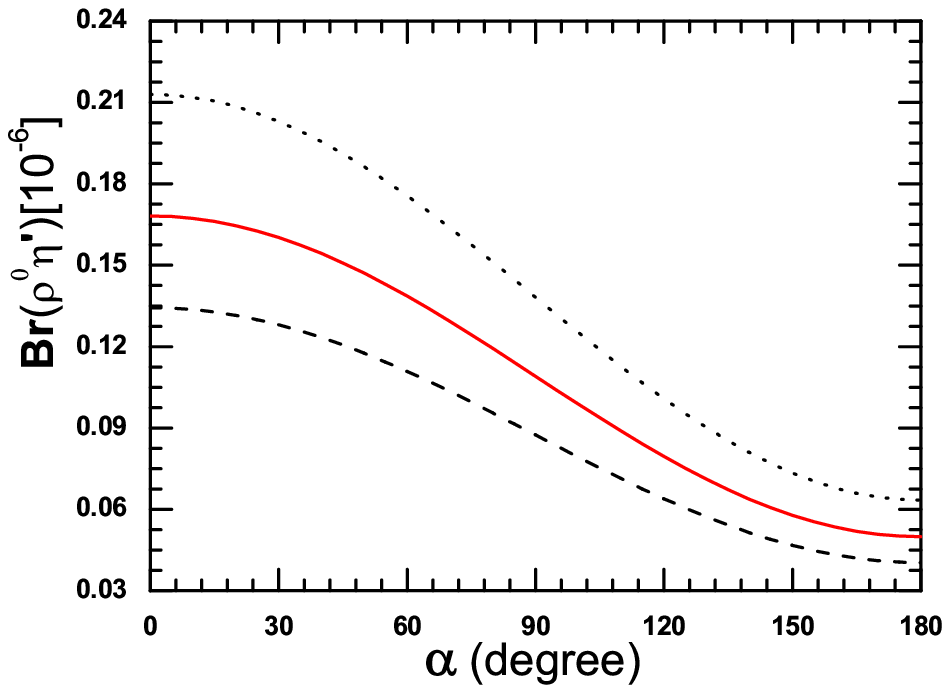}
\vspace{0.3cm}
\caption{The $\alpha$ dependence of the branching
ratios (in units of $10^{-6}$) of $B^0\to\rho^0\etap$ decays for
$\omega_b=0.36$ GeV (dotted curve), 0.40 GeV (solid curve) and 0.44
GeV (dashed curve). }\label{fig:ro0bralf}
\end{center}
\end{figure}

\begin{figure}[tb]
\begin{center}
\includegraphics[scale=0.7]{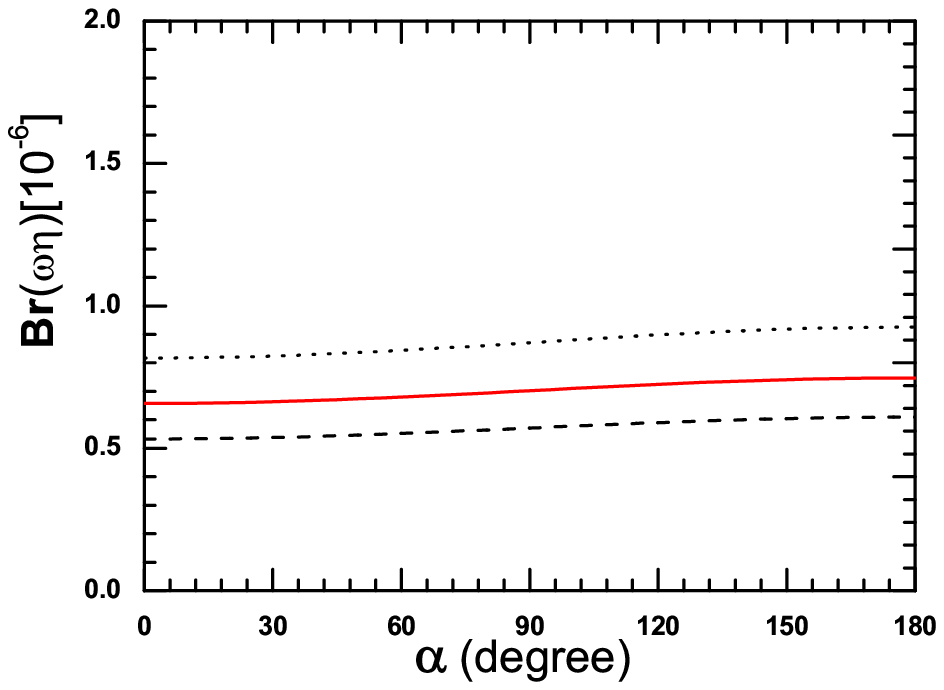}
\includegraphics[scale=0.7]{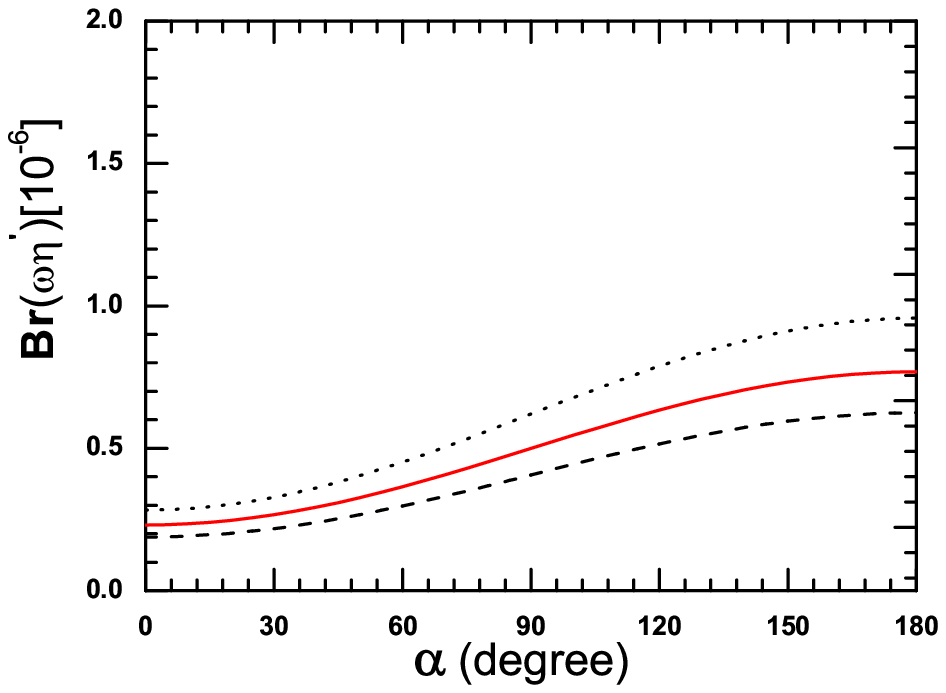}
\vspace{0.3cm}
\caption{The $\alpha$ dependence of the branching
ratios (in units of $10^{-6}$) of $B^0\to\omega\etap$ decays for
$\omega_b=0.36$ GeV (dotted curve), 0.40 GeV (solid curve) and 0.44
GeV (dashed curve). }\label{fig:omigbralf}
\end{center}
\end{figure}

In Figs.~\ref{fig:rozbralf}, \ref{fig:ro0bralf} and
\ref{fig:omigbralf} we show the $\alpha$ and $\omega_b$-dependence of the pQCD
predictions for the branching ratios of $B\to\rho\etap$,
$B\to\omega\etap$ decays for $\omega_b=0.4\pm0.04$ GeV, and
$\alpha=[0^\circ,180^\circ]$, $a_2=0.115$  and $a_{2\rho}=a_{2\omega}=0.15$.

From the numerical results and the figures, we observe that
\begin{itemize}
\item
For $B^\pm \to \rho^\pm \eta$ decay, the inclusion of the considered NLO corrections
can improve the agreement between the pQCD prediction and the data. But for
$B^\pm \to \rho^\pm \etar$ decay, we are not so lucky.
Although the pQCD predictions for $Br(B\pm \to \rho^\pm \etap)$ agree with the data
within one standard deviation, but the predicted pattern of $Br(B\pm \to \rho^\pm \eta)
> Br(B^\pm \etar)$ in both the pQCD and QCDF is contrary to the observed one.

\item
For $B^0 \to \rho^0(\omega,\phi) \etap$ decays, the pQCD predictions for their Br's
are consistent with currently  available upper limits.
Except for $Br(B\to \phi\etar)$, the inclusion of the partial NLO contributions
to other decays can enhance their Br's by a factor of two to ten, and generally larger than
the QCDF predictions, which will be tested by the
forthcoming LHCb experiment.

\end{itemize}

\subsection{CP-violating asymmetries }

   Now we turn to the evaluations of the CP-violating asymmetries of
$B \to \rho(\omega,\phi) \etap$ decays in pQCD approach. For $B^+ \to
\rho^+ \eta$ and $B^+ \to \rho^+ \eta^\prime$ decays, the direct
CP-violating asymmetries $\acp$ can be defined as:
 \beq
\acp^{dir} =  \frac{|\overline{\cal M}_f|^2 - |{\cal M}_f|^2}{
 |\overline{\cal M}_f|^2+|{\cal M}_f|^2}, \label{eq:acp1}
 \eeq
The pQCD predictions for the direct CP-violating
asymmetries of $B^\pm \to \rho^\pm \etap$ decays are listed in Table \ref{table2}.
For comparison, we also list currently available experimental results
\cite{pdg2006,hfag} and the numerical results evaluated in
the framework of the QCD factorization (QCDF) \cite{npb675}.

\begin{table}[thb]
\begin{center}
\caption{ The pQCD predictions for the direct CP-violating asymmetries
of $B^\pm \to \rho^\pm \etap$ decays (in units of $10^{-2}$).}
\label{table2}
\vspace{0.2cm}
\begin{tabular}{l  | c c c c |c c c } \hline \hline
 Mode&  LO &  +VC & +QL & +MP & NLO &Data &QCDF\\
\hline
$\acp^{dir}(B^\pm \to \rho^\pm \eta)$ &0.0 &1.3&1.4&-0.1&1.9&$1.0\pm16.0$&2.4\\
$\acp^{dir}(B^\pm \to \rho^\pm \eta^{\prime})$ &-6.8&-25.3&-5.7&-7.1&-25.0&$-4.0\pm28$&-4.1
\\
 \hline \hline
\end{tabular}
\end{center}\end{table}

\begin{figure}[tb]
\begin{center}
\includegraphics[scale=0.7]{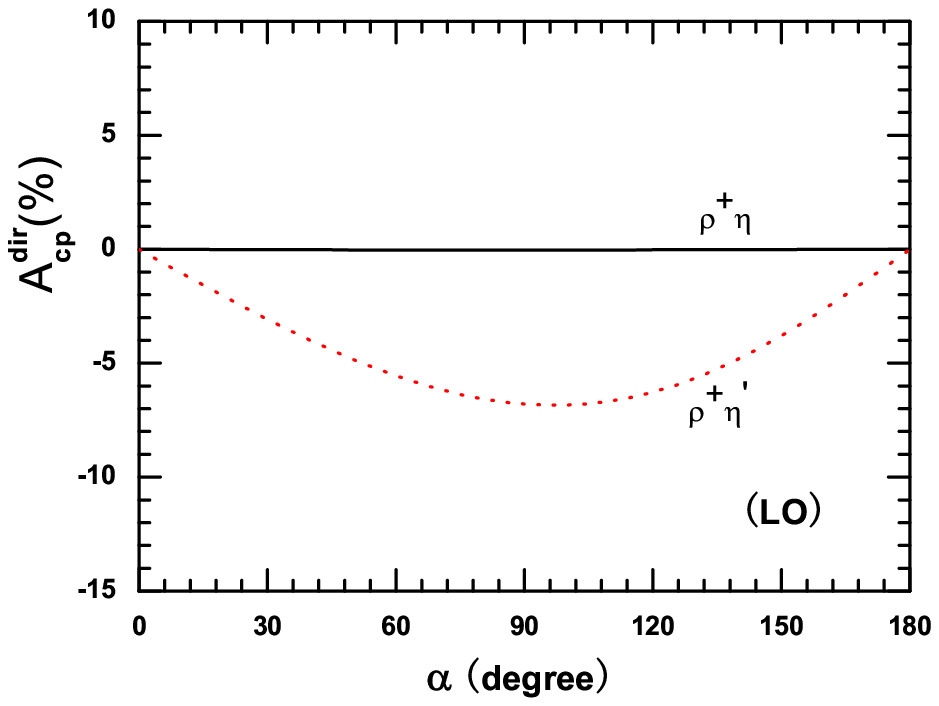}
\includegraphics[scale=0.7]{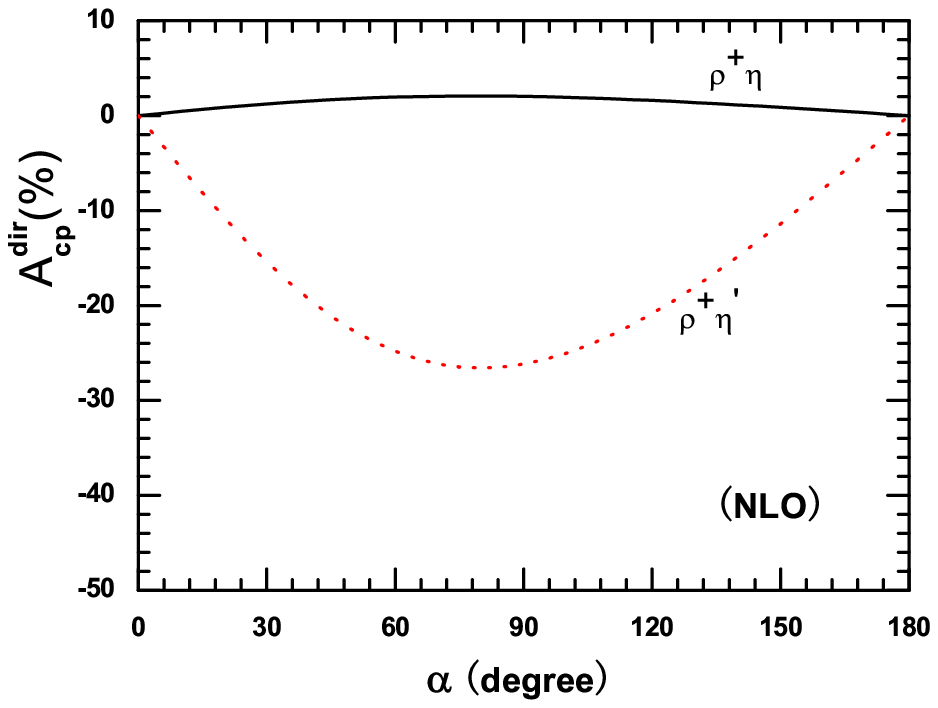}
\vspace{0.3cm}
\caption{The $\alpha$ an $\omega_b$-dependence of the CP-violating asymmetries
of $B^\pm \to\rho^\pm \etap$ decays for $\alpha=[0^\circ,180^\circ]$ and
$\omega_b=0.36$ GeV (dotted curve), 0.40 GeV (solid curve) and 0.44
GeV (dashed curve). }\label{fig:cpv1}
\end{center}
\end{figure}

The NLO pQCD predictions for the central values of the direct CP-violating
asymmetries  and the major theoretical errors for $B^\pm  \to \rho^\pm  \etap$ decays are
\beq
\acp^{dir}(B^\pm \to \rho^\pm \eta) &=& \left [1.9
^{+0.1}_{-0.0}(\omega_b)^{+0.2}_{-0.3}(\alpha)^{+0.1}_{-0.0}(a_{2})
^{+0.6}_{-0.5}(a_{2\rho})\right ]
 \times 10^{-2} \label{eq:acp-a},\quad \\
\acp^{dir}(B^\pm \to \rho^\pm \eta^\prime) &=& \left [-25.0
^{+0.4}_{-0.3}(\omega_b)^{+4.1}_{-1.6}(\alpha)^{+0.8}_{-0.7}(a_{2})
^{+2.1}_{-1.8}(a_{2\rho})\right ]  \times 10^{-2} \label{eq:acp-b},
\eeq
where the major theoretical errors come from the variations of
$\omega_b=0.4\pm 0.04$ GeV, $\alpha=100^\circ \pm 20^\circ$,
Gegenbauer coefficients $a_2=0.115\pm0.115$, $a_{2\rho}=a_{2\omega}=
0.15\pm0.15$. Both the pQCD and QCDF predictions are consistent with the data because
of the still large theoretical and experimental errors. In Fig.~\ref{fig:cpv1}, one shows the
$\alpha$ and $\omega_b$-dependence of the LO and NLO pQCD predictions for the
CP-violating asymmetries of $B^\pm \to \rho^\pm \etap$.

As to the CP-violating asymmetries for the neutral decays $B^0 \to
\rho^0(\omega) \etap$, the effects of $B^0-\bar{B}^0$ mixing should
be considered. The CP-violating asymmetries for such decays
are  time dependent and can be defined as
\beq
A_{CP} &\equiv& \frac{\Gamma\left (\overline{B_d^0}(\Delta t) \to
f_{CP}\right) - \Gamma\left(B_d^0(\Delta t) \to f_{CP}\right )}{
\Gamma\left (\overline{B_d^0}(\Delta t) \to f_{CP}\right ) +
\Gamma\left (B_d^0(\Delta t) \to f_{CP}\right ) }\non &=&
A_{CP}^{dir} \cos (\Delta m  \Delta t) + A_{CP}^{mix} \sin (\Delta m
\Delta t), \label{eq:acp-def}
\eeq
where the direct and mixing induced CP-violating
asymmetries $A_{CP}^{dir}$ and $A_{CP}^{mix}$ can be written as
\beq
\acp^{dir}=\frac{ \left | \lambda_{CP}\right |^2 -1 }
{1+|\lambda_{CP}|^2}, \qquad A_{CP}^{mix}=\frac{ 2Im
(\lambda_{CP})}{1+|\lambda_{CP}|^2}, \label{eq:acp-dm}
\eeq
with the CP-violating parameter $\lambda_{CP}$ is
\beq
\lambda_{CP} = - \frac{ V_{tb}^*V_{td} \langle V \etap |H_{eff}|
\overline{B}^0\rangle} { V_{tb}V_{td}^* \langle V \etap
|H_{eff}| B^0\rangle}, \label{eq:lambda2}
\eeq

If we integrate the time variable $t$, we will get the total CP
asymmetries for $B^0 \to V \etap$ decays,
\beq
A_{CP}=\frac{1}{1+x^2} A_{CP}^{dir} + \frac{x}{1+x^2} A_{CP}^{mix},
\eeq
where $x=\Delta m/\Gamma=0.775$ for the $B^0-\overline{B}^0$
mixing \cite{pdg2006}.

\begin{table}[tb]
\begin{center}
\caption{ The pQCD predictions for the direct, mixing-induced and total CP
asymmertries of $B^0 \to \rho^0(\omega) \etap$ decays (in unit of $10^{-2}$).}
\label{table3}
\vspace{0.2cm}
\begin{tabular}{l | r  r r r r | r r} \hline \hline
 Mode&  LO & $LO_{{\rm NLOWC}}$ & +VC & +QL & +MP & NLO& QCDF \\
\hline
$\acp^{dir}(B^0 \to \rho^0 \eta)$  &$ -78.3$ &$  -94.4$ &$  -91.4$ &$-79.0$ &$-97.4 $ &$-89.6$  &$  -  $ \\
$\acp^{dir}(B^0 \to \rho^0 \etar)$ &$ 77.3 $ &$  -42.6$ &$  -79.8$ &$-96.8$ &$-86.2 $ &$-75.7$  &$  -  $ \\
$\acp^{dir}(B^0 \to \omega \eta)$  &$ 94.6 $ &$  53.4 $ &$  45.7 $ &$48.7 $ &$32.1  $ &$33.5 $  &$  33.4$  \\
$\acp^{dir}(B^0 \to \omega \etar)$ &$ 30.0 $ &$  26.9 $ &$  26.5 $ &$15.4 $ &$-12.1 $ &$16.0 $  &$  -0.2$  \\ \hline
$\acp^{mix}(B^0 \to \rho^0 \eta)$ &$ 44.7 $&$ 26.1 $&$ 16.9 $&$ 14.1$&$ 17.5$&$ 22.7$&$- $\\
$\acp^{mix}(B^0 \to \rho^0\etar)$ &$-24.0 $&$ -15.9$&$ -46.0$&$ -3.5$&$-14.4$&$-49.0$&$- $ \\
$\acp^{mix}(B^0 \to \omega \eta)$ &$-7.4  $&$ 83.8 $&$ 40.3 $&$ 81.5$&$ 80.2$&$ 39.0$&$- $ \\
$\acp^{mix}(B^0 \to \omega\etar)$ &$ 58.9 $&$ 88.7 $&$ 78.6 $&$ 82.2$&$71.6 $&$77.0 $&$- $ \\ \hline
$\acp^{tot}(B^0 \to \rho^0 \eta) $ &$ -24.0$&$ -46.4$&$ -48.9 $&$ -42.5$&$-52.4$&$-45.0$&$- $ \\
$\acp^{tot}(B^0 \to \rho^0 \etar)$ &$ 35.5 $&$-34.3 $&$ -72.1 $&$ -62.2$&$-60.8$&$-71.0$&$- $ \\
$\acp^{tot}(B^0 \to \omega \eta) $ &$ 54.2 $&$ 73.9 $&$ 48.1  $&$ 69.9 $&$58.9 $&$39.8 $&$- $   \\
$\acp^{tot}(B^0 \to \omega \etar)$ &$ 48.4 $&$59.7  $&$ 54.6  $&$ 49.4 $&$27.1 $&$47.3 $&$- $ \\
 \hline \hline
\end{tabular}
\end{center}\end{table}

The pQCD predictions for the CP-violating asymmetries and the total CP violation
when different NLO contributions are included step by step are listed in Table III.
The pQCD predictions with major theoretical errors are given in
Eqs.~(\ref{eq:acp-f}-\ref{eq:tot-b}):
\beq
\acp^{dir}(B^0 \to \rho^0 \eta) &=& \left [-89.6 ^{+1.9}_{-0.9}(\omega_b)^{+13.7}_{-3.9}(\alpha)^{+0.7}_{-0.1}(a_{2})
^{+4.6}_{-9.0}(a_{2\rho})\right ]  \times 10^{-2} \label{eq:acp-c},\quad \non
\acp^{dir}(B^0 \to \rho^0 \eta^\prime) &=& \left [-75.7
^{+5.6}_{-4.8}(\omega_b)^{+13.1}_{-7.0}(\alpha)^{+6.3}_{-4.0}(a_{2})
^{+12.9}_{-9.9}(a_{2\rho})\right ]  \times 10^{-2} \label{eq:acp-d},\non
\acp^{dir}(B^0 \to \omega \eta)&=&\left[33.5 ^{+1.0}_{-1.4}(\omega_b)_{-4.6}^{+0.8}(\alpha)
_{-6.8}^{+5.9}(a_2)_{-4.4}^{+3.9}(a_{2\omega}) \right ] \times 10^{-2},\label{eq:acp-e}\non
\acp^{dir}(B^0 \to \omega \eta^{\prime})&=&\left[16.0
^{+0.1}_{-0.9}(\omega_b)^{+3.3}_{-3.9}(\alpha) ^{+2.2}_{-3.2}(a_2)
^{+1.7}_{-2.0}(a_{2\omega})\right ] \times 10^{-2}, \label{eq:acp-f}
\eeq
\beq
\acp^{mix}(B^0 \to \rho^0 \eta) &=& \left [22.7
\pm6.1(\omega_b)^{+13.9}_{-21.8}(\alpha)^{+9.6}_{-12.5}(a_{2})
^{+23.6}_{-26.5}(a_{2\rho})\right ]
 \times 10^{-2} \label{eq:mix-a}, \non
\acp^{mix}(B^0 \to \rho^0 \eta^\prime) &=&\left [-49.0
^{+1.9}_{-0.8}(\omega_b)^{+16.0}_{-8.1}(\alpha)
_{-4.2}^{+1.8}(a_2)^{+18.6}_{-17.8}(a_{2\rho}) \right ]
 \times 10^{-2} \label{eq:mix-b}, \non
\acp^{mix}(B^0 \to \omega \eta) &=& \left [39.0
^{+0.3}_{-0.2}(\omega_b)^{+50.6}_{-66.2}(\alpha)^{+5.9}_{-3.3}(a_{2})
^{+2.9}_{-1.9}(a_{2\omega})\right ]
 \times 10^{-2} \label{eq:mix-a3}, \non
\acp^{mix}(B^0 \to \omega \eta^\prime) &=&\left [77.0
^{+0.4}_{-0.1}(\omega_b)^{+22.0}_{-52.9}(\alpha)
_{-0.1}^{+0.9}(a_2)^{+0.3}_{-0.0}(a_{2\omega}) \right ]
 \times 10^{-2} \label{eq:mix-b3},
 \eeq
 \beq
\acp^{tot}(B^0 \to \rho^0 \eta) &=& \left
[-45.0_{-1.7}^{+2.4}(\omega_b)^{+15.3}_{-13.0}(\alpha)
_{-6.1}^{+5.1}(a_2)^{+17.1}_{-15.7}(a_{2\rho}) \right ]
 \times 10^{-2} \label{eq:tot-a}, \non
\acp^{tot}(B^0 \to \rho^0 \eta^\prime) &=& \left [-71.0
^{+2.9}_{-2.1}(\omega_b)^{+4.2}_{-0.0}(\alpha)
^{+1.9}_{-1.7}(a_2)_{-0.6}^{+2.8}(a_{2\rho}) \right ]
 \times 10^{-2} \label{eq:tot-b},\non
 \acp^{tot}(B^0 \to \omega \eta) &=& \left
[39.8_{+0.6}^{-0.7}(\omega_b)^{+20.2}_{-31.5}(\alpha)
_{-5.8}^{+6.6}(a_2)^{+3.9}_{-3.6}(a_{2\omega}) \right ]
 \times 10^{-2} \label{eq:tot-a3}, \non
\acp^{tot}(B^0 \to \omega \eta^\prime) &=& \left [47.3
^{+0.2}_{-0.6}(\omega_b)^{+8.2}_{-23.6}(\alpha)
_{-2.1}^{+1.8}(a_2)\pm1.2(a_{2\omega}) \right ]
 \times 10^{-2} \label{eq:tot-b3},
 \eeq
where the dominant errors come from the variations of
$\omega_b=0.4\pm 0.04$ GeV, $\alpha=100^\circ\pm20^\circ$, and
Gegenbauer coefficient $a_2=0.115\pm0.115$, $a_{2\rho}=a_{2\omega}= 0.15\pm0.15$.

For the CP-violating asymmetries of $B^0\to \rho^0(\omega) \etap$ decays, unfortunately,
there is no data available currently.
For $B^0 \to \phi \etap$ decays, there is no CP violation. The reasons are simple: (a) the total
decay amplitude at the LO level as given in Eq.~(\ref{eq:phi}) is proportional to only
one CKM factor $\xi_t$; and (b) among the NLO contributions considered here,
only the vertex correction ( real correction ) is relevant for this decay mode.

%%%%%%%%%%%%%%%%%%%%%%%%%%%%%%%%%%%%%%%%%%%%%%%%%%%%%%%%%%%%%%%%%
\section{summary }

In this paper,  we calculate some NLO contributions to the branching ratios and CP-violating
asymmetries of $B^\pm \to \rho^\pm \etap$ and $B^0 \to \rho^0(\omega,\phi) \etap$
decays by employing the pQCD factorization approach.

From our calculations and phenomenological analysis, we found the following results:
\begin{itemize}

\item
The pQCD predictions for the form factors of $B\to \rho, \omega$ and $\etap$ transitions are
$A_0^{B\to \rho}(0)= 0.32_{-0.04}^{+0.05}(\omega_b)$,
$A_0^{B\to \omega}(0)= 0.29_{-0.03}^{+0.04}(\omega_b)$ and
$F^{B\to\etap}_0(0)=0.22\pm{0.03}(\omega_b)$ for $\omega_b=0.40\pm0.04$GeV,
which agree very well with those obtained in QCD sum rule calculations.

\item
For $B^\pm \to \rho^\pm \eta$ decay, the inclusion of partial NLO contributions can improve
the agreement between the pQCD predictions and the measured values.
For the neutral decays, the NLO contributions can provide significant
enhancements to the LO predictions:
\beq
Br(\ B^\pm \to \rho^\pm \eta) &=& \left [6.7 ^{+2.6}_{-1.9}\right ] \times 10^{-6}
 ,\non
Br(\ B^\pm \to \rho^\pm \eta^{\prime}) &=&\left [4.6 ^{+1.6}_{-1.4}\right ] \times 10^{-6}
,\non
Br(\ B^0 \to \rho^0 \eta) &=& \left [1.3 ^{+1.3}_{-0.6}\right ] \times 10^{-7}
,\non
Br(\ B^0 \to \rho^0 \eta^{\prime}) &=& \left [1.0 \pm 0.5 \right ] \times 10^{-7}
,\non
Br(\ B^0 \to \omega \eta) &=& \left [7.1
^{+3.7}_{-2.8}\right ] \times 10^{-7} ,\non
Br(\ B^0 \to \omega \eta^{\prime}) &=& \left [5.5 ^{+3.1}_{-2.6} \right ]
\times 10^{-7} ,\non
Br(\ B^0 \to \phi \eta) &=& \left [1.1 ^{+6.2}_{-0.9} \right ] \times 10^{-8}
,\non
Br(\ B^0 \to \phi \eta^{\prime}) &=& \left [1.7 ^{+16.1}_{-1.0}
\right ] \times 10^{-8} \label{eq:br0-etap8},
\eeq
where the various errors as given  in Eq.~(\ref{eq:etap11}) have been added in quadrature.

\item
The pQCD predictions for $\acp^{dir}(B^\pm \to \rho^\pm \etap)$ are consistent
with the data, but both the  theoretical and experimental errors are still large.
For other neutral decays, the pQCD predictions for CP violating asymmetries are generally large
in magnitude and could be tested by the forthcoming LHCb experiments.

\item
Only the NLO contributions from vertex correction, quark-loops and chromo-magnetic penguins
are calculated here. The NLO corrections from the hard-spectator and annihilations diagrams
are still absent now. It is an urgent task to do the relevant calculations, in order to
provide a complete NLO calculation in the pQCD approach.

\end{itemize}

\begin{acknowledgments}

The authors are very grateful to Kavli Institute for Theoretical Physics China, Beijing, China,
where part of this work was done.
This work is partly supported  by the National Natural Science
Foundation of China under Grant No.10575052 and 10735080.

\end{acknowledgments}

%%%%%%%%%%%%%%%%%%%%%%%%%%%%%%%%%%%%%%%%%%%%%%%%%%%%%%%%%%%%%%%%%%%%%%%%%%%%%%%%%%
%                                        Appendix
%%%%%%%%%%%%%%%%%%%%%%%%%%%%%%%%%%%%%%%%%%%%%%%%%%%%%%%%%%%%%%%%%%%%%%%%%%%%%%%%5

\begin{appendix}

\section{Wave Functions and Input Parameters}\label{sec:wf2}

The B meson is treated as a heavy-light system.
For the B meson wave function, since the contribution of $\ov{\phi}_B$ is
numerically small ~\cite{luyang}, we here only consider the
contribution of Lorentz structure
\beq
\Phi_B= \frac{1}{\sqrt{2N_c}}
(\psl_B +m_B) \gamma_5 \phi_B ({\bf k_1}), \label{bmeson}
\eeq
with
\beq
\phi_B(x,b)&=& N_B x^2(1-x)^2 \mathrm{exp} \left
 [ -\frac{M_B^2\ x^2}{2 \omega_{b}^2} -\frac{1}{2} (\omega_{b} b)^2\right],
 \label{phib}
\eeq
where $\omega_{b}$ is a free parameter and we take
$\omega_{b}=0.4\pm 0.04$ GeV in numerical calculations, and
$N_B=101.445$ is the normalization factor for $\omega_{b}=0.4$.

For the considered decays, the vector meson V is longitudinally
polarized. The longitudinal polarized component of the wave function
is defined as:
\beq
\phi_V=\frac{1}{\sqrt{2N_c}}\left\{\esl\left[m_V\phi_V(x)+\pvsl\phi_V^t(x)\right]
+m_V\phi^s_V(x)\right\},
\eeq
where the first term is the leading twist (twist-2) wave function,
while the second and third terms are twist-3 wave functions.

The twist-2 DA's for longitudinally polarized vector meson can be
parameterized as:
\beq
\pva(x) &=&  \frac{f_V}{2\sqrt{2N_c} }    6x (1-x)
    \left[1+a_{2V}C^{3/2}_2(2x-1)\right],\label{piw1}
 \end{eqnarray}
for $V= \rho, \omega, \phi$; and $f_V$ is the decay constant of the
vector meson with longitudinal polarization, and numerically
\cite{pgr}:
\beq
f_{\rho}=216 MeV,\quad f_{\omega}=187 MeV, \quad f_{\phi}=215 MeV.
\eeq

The Gegenbauer coefficients have been studied extensively in the
literature. Here we adopt the following values from the recent
updates \cite{pgr}:
\beq
a_{2\rho}=a_{2\omega}=0.15\pm0.15, \quad
a_{2\phi}=0.2\pm0.2.
\eeq
We shall vary the Gegenbauer coefficients
of the twist-2 distribution amplituds by $100\%$, which is larger
than the error specified in \cite{pgr}. Therefore, the theoretical
uncertainty of our predictions from this source is conservative.

As for the twist-3 DAs $\pvp$ and $\pvt$, we adopt their asymptotic form \cite{likstar}:
\beq
 \pvp(x) =   \frac{3f^T_V}{2\sqrt{2N_c}}(1-2x),\quad \pvt(x)
 =\frac{3f^T_V}{2\sqrt{2N_c}}(2x-1)^2,\label{piw3}
 \eeq

For $\etap$ meson, the wave function for $q\bar{q}$ ($q=u,d$) components
of $\etap$ meson are given as
\beq
\Phi_{\eta_q}(P,x,\zeta)\equiv \frac{1}{\sqrt{2N_C}} \gamma_5 \left [ \psl
\phi_{\eta_q}^{A}(x)+m_0^{\eta_q} \phi_{\eta_q}^{P}(x)+\zeta m_0^{\eta_q} ( \vsl
\nsl - v\cdot n)\phi_{\eta_q}^{T}(x) \right ],
\eeq
where $P$ and $x$ are the momentum and the momentum fraction of
$\eta_q$, respectively. We assumed here that the wave
function of $\eta_q$ is same as the $\pi$ wave function.
The parameter $\zeta$ is either $+1$ or $-1$ depending on the
assignment of the momentum fraction $x$. The $s\bar s$ component of
the wave function can be similarly defined.

%%%%--------------------------------------------------------------

For the mixing of $\eta-\etar$ system, we here use the
quark-flavor basis, that is the $\eta_q = (u\bar u +d\bar d)/\sqrt{2} $ and $\eta_s=s\bar s$.
Then the physical states $\eta$ and $\etar$ are related to the flavor states through a
single mixing angle $\phi$,
\beq
\left(\begin{array}{c} \eta \\ \eta^{\prime} \end{array} \right)
=\left(\begin{array}{cc}
 \cos{\phi} & -\sin{\phi} \\
 \sin{\phi} & \cos{\phi} \\ \end{array} \right)
 \left(\begin{array}{c} \eta_q \\ \eta_s \end{array} \right),
\label{eq:e-ep}
\eeq
The relation between the decay constants $(f_\eta^q, f_\eta^s,f_{\etar}^q,f_{\etar}^s)$ and
$(f_q,f_s,)$ can be written as
\beq
f^q_\eta &=&  f_q cos\phi, \quad  f^s_{\eta} = -f_s \sin\phi, \non
f^q_{\etar}  &=& f_q \sin\phi, \quad  f^s_{\etar} = f_s \cos\phi.
 \label{eq:op}
\eeq

 The chiral enhancement $m_0^q$ and $m_0^s$ associated
 with the two-parton twist-3 $\eta_q$ and $\eta_s$ meson
 distribution amplitudes have been defined as \cite{nlo05}
 \beq
 m^{q}_{0}&=&\frac{m^2_{qq}}{2m_q}=\frac{1}{2m_q}[m^2_{\eta}\cos^2\phi
 +m^2_{\eta^{\prime}}\sin^2\phi-\frac{\sqrt{2}f_s}{f_q}(m^2_{\eta^{\prime}}-m^2_{\eta})
 \cos\phi\sin\phi],\label{eq:m0q}\\
m^{s}_{0}&=&\frac{m^2_{ss}}{2m_s}=\frac{1}{2m_s}[m^2_{\eta^{\prime}}\cos^2\phi
 +m^2_{\eta}\sin^2\phi-\frac{\sqrt{2}f_q}{f_s}(m^2_{\eta^{\prime}}-m^2_{\eta})
 \cos\phi\sin\phi],\label{eq:m0s}
 \eeq
by assuming the exact isospin symmetry $m_q=m_u=m_d$.
The three input parameters $f_q,f_s$ and $\phi$ have been
extracted from the data of the relevant exclusive processes\cite{feldmann}:
\beq
f_q=(1.07\pm 0.02)f_{\pi},\quad f_s=(1.34\pm 0.06)f_{\pi},\
\quad \phi=39.3^\circ\pm 1.0^\circ,
\eeq

It is still unclear for the possible gluonic component of $\etar$ meson.
From currently known studies\cite{liu06,guodq07,guo07}  we believe that
there is no large room left for the contribution due to the gluonic
component of $\etar$, and therefore will neglect the possible
gluonic component in $\etar$ mson.

%%============================================

The distribution amplitude $\phi_{\eta_q}^{A,P,T}$ represents the axial vector, pseudoscalar
and tensor component of the wave function respectively \cite{ball98}.
They are given as:
\begin{eqnarray}
 \phi_{\eta_q}^A(x) &=&  \frac{f_{ \eta_q }}{2\sqrt{2N_c} }
    6x (1-x)
    \left[1+a^{\eta_q}_1C^{3/2}_1(2x-1)+a^{\eta_q}_2 C^{3/2}_2(2x-1)
    \right.\non && \left.+a^{\eta_q}_4C^{3/2}_4(2x-1)
  \right],\label{piw11}\\
 \phi_{\eta_q }^P(x) &=&   \frac{f_{\eta_q }}{2\sqrt{2N_c} }
   \left[ 1+(30\eta_3-\frac{5}{2}\rho^2_{\eta_q } )C^{1/2}_2(2x-1)
\right.\non && \left.
   -3\left\{\eta_3\omega_3+\frac{9}{20}\rho^2_{\eta_q }
   (1+6a^{\eta_q }_2)\right\} C^{1/2}_4(2x-1)\right]  ,\\
 \phi_{\eta_q}^T(x) &=&  \frac{f_{\eta_q }}{2\sqrt{2N_c} } (1-2x)
   \left[ 1+6\left (5\eta_3-\frac{1}{2}\eta_3\omega_3
   -\frac{7}{20}\rho^2_{\eta_q}-\frac{3}{5}\rho^2_{\eta_q }a_2^{\eta_q}\right )
   \right. \non && \left.
   \cdot \left (1-10x+10x^2 \right )\right] ,\quad\quad\label{piw4}
 \end{eqnarray}
with
\beq
\rho_{\eta_q}=2m_q/m_{qq},\quad
a^{\eta_{q\bar{q}}}_1&=&0,\quad
a^{\eta_{q\bar{q}}}_2=0.115\pm0.115,\quad a^{\eta_{q\bar{q}}}_4=-0.015.
\eeq
and the Gegenbauer polynomials $C^{\nu}_n(t)$,
\beq
C^{1/2}_2(t)&=&\frac{1}{2}(3t^2-1), \qquad C^{1/2}_4(t)=\frac{1}{8}(3-30t^2+35t^4), \\
C^{3/2}_1(t)&=&3t, \qquad C^{3/2}_2(t)=\frac{3}{2}(5t^2-1),\\
C^{3/2}_4(t)&=&\frac{15}{8}(1-14t^2+21t^4).\label{eq:c124}
\eeq
The Gegenbauer coefficients can vary by $100\%$, but we do not consider
the uncertainty from the coefficients $a^{\eta_{q\bar{q}}}_4$, to
which our predictions are insensitive. The values of other
parameters are $\eta_3=0.015$ and $\omega=-3.0$.

As to the wave function of the $s\bar{s}$ components, we also use
the same form as $q\bar{q}$ but with some parameters changed :
\beq
\rho_{\eta_s}=2m_s/m_{ss},\quad
a^{\eta_s}_1=0, \quad
a^{\eta_s}_2=0.115\pm0.115,\quad
a^{\eta_s}_4=-0.015.
\eeq

Besides those specified in the text, the following input parameters will also be used
in the numerical calculations:
\beq
f_\pi &=& 130 {\rm MeV}, \quad  f_B = 210 {\rm MeV},\quad
m_{\eta}=547.5{\rm MeV},\quad m_{\eta^{\prime}}=957.8{\rm MeV},\non
m_{q}&=&5.6{\rm MeV},\quad m_{s}= 130\pm 30 {\rm MeV},\quad m_B =5.28 {\rm GeV},  \non
m_{\rho}&=&774{\rm MeV}, \quad m_{\omega}=780{\rm MeV}\quad
m_{\phi}=1.02{\rm GeV}, \quad m_W = 80.41{\rm GeV}, \non
\tau_{B^0} &=& 1.528 {\rm ps}, \quad \tau_{B^+} = 1.643 {\rm ps},
\label{para}
\eeq

For the CKM quark-mixing matrix, we use the Wolfenstein parametrization as given
in Ref.\cite{pdg2006,hfag}.
\beq
V_{ud}&=&0.9745, \quad V_{us}=\lambda = 0.2200, \quad |V_{ub}|=4.31\times 10^{-3},\non
V_{cd}&=&-0.224, \quad  V_{cd}=0.996, \quad V_{cb}=0.0413, \non
|V_{td}|&=& 7.4 \times 10^{-3}, \quad V_{ts}=-0.042, \quad |V_{tb}|=0.9991,
\label{eq:vckm}
\eeq
with the CKM angles $\beta=21.6^\circ$, $\gamma =60^\circ \pm 20^\circ $ and
$\alpha=100^\circ \pm 20^\circ $.

%%=================================

\section{Related Functions }\label{sec:aa}

 We show here the function $h_i$'s, coming from the Fourier transformations  of $H^{(0)}$,
 \beq
 h_e(x_1,x_2,b_1,b_2)&=&  K_{0}\left(\sqrt{x_1 x_2} m_B b_1\right)
\left[\theta(b_1-b_2)K_0\left(\sqrt{x_2} m_B
b_1\right)I_0\left(\sqrt{x_2} m_B b_2\right)\right. \non & &\;\left.
+\theta(b_2-b_1)K_0\left(\sqrt{x_2}  m_B b_2\right)
I_0\left(\sqrt{x_2}  m_B b_1\right)\right] S_t(x_2), \label{he1}
\eeq \beq h_a(x_2,x_3,b_2,b_3)&=& K_{0}\left(i\sqrt{(1-x_2)x_3} m_B
b_2\right)
 \left[\theta(b_3-b_2)K_0\left(i \sqrt{x_3} m_B
b_3\right)I_0\left(i \sqrt{x_3} m_B b_2\right)\right. \non &
&\;\;\;\;\left. +\theta(b_2-b_3)K_0\left(i \sqrt{x_3}  m_B
b_2\right) I_0\left(i \sqrt{x_3}  m_B b_3\right)\right] S_t(x_3),
\label{eq:he3}
\eeq
 \beq
 h_{n}(x_1,x_2,x_3,b_1,b_3) &=& \biggl\{\theta(b_1-b_3)\mathrm{K}_0(M_B\sqrt{x_1 x_2} b_1)
  \mathrm{I}_0(M_B\sqrt{x_1 x_2} b_3)\non &+ & \theta(b_3-b_1)\mathrm{K}_0(M_B\sqrt{x_1 x_2} b_3)
  \mathrm{I}_0(M_B\sqrt{x_1 x_2} b_1) \biggr\} \non && \cdot\left(
\begin{matrix}
 \frac{\pi i}{2}\mathrm{H}_0(\sqrt{(x_2(x_3-x_1))} M_B b_3), & \text{for}\quad x_1-x_3<0 \\
 \mathrm{K}_0^{(1)}(\sqrt{(x_2(x_1-x_3)}M_B b_3), &
 \text{for} \quad x_1-x_3>0
\end{matrix}\right),
\label{eq:pp1} \eeq \beq h_{na}(x_1,x_2,x_3,b_1,b_3) &=&
\biggl\{\theta(b_1-b_3) \mathrm{K}_0(i \sqrt{(1-x_2) x_3} b_1 M_B)
 \mathrm{I}_0(i \sqrt{(1-x_2) x_3} b_3 M_B)\non &+&(\theta(b_3-b_1) \mathrm{K}_0(i \sqrt{(1-x_2) x_3} b_3 M_B)
 \mathrm{I}_0(i \sqrt{(1-x_2) x_3} b_1 M_B) \biggr\}
 \non
& & \cdot\left(
\begin{matrix}
 \mathrm{K}_0(M_B\sqrt{(x_1-x_3)(1-x_2)}b_1), & \text{for}\quad x_1-x_3>0 \\
 \frac{\pi i}{2} \mathrm{H}_0^{(1)}(M_B\sqrt{(x_3-x_1)(1-x_2)}b_1), &
 \text{for} \quad x_1-x_3<0
\end{matrix} \right),
\label{eq:pp4}
\eeq
 \beq
 h_{na}^\prime(x_1,x_2,x_3,b_1,b_2) &=&
 \biggl\{\theta(b_1-b_3) \mathrm{K}_0(i \sqrt{(1-x_2) x_3} b_1 M_B)
 \mathrm{I}_0(i \sqrt{(1-x_2) x_3} b_3 M_B)
 \non
&+& \theta(b_3-b_1) \mathrm{K}_0(i \sqrt{(1-x_2) x_3} b_3
M_B)\mathrm{I}_0(i \sqrt{(1-x_2) x_3} b_1 M_B)\biggr\} \non &&\cdot
\left(
\begin{matrix}
 \mathrm{K}_0(M_B F_{1}b_1), & \text{for}\quad F_{1}^2>0 \\
 \frac{\pi i}{2} \mathrm{H}_0^{(1)}(M_B \sqrt{|F_{1}^2|}b_1), &
 \text{for}\quad F_{1}^2<0
\end{matrix}\right), \label{eq:pp3}
\eeq  where $J_0$ is the Bessel function and $K_0$, $I_0$ are
modified Bessel functions $K_0 (-i x) = -(\pi/2) Y_0 (x) + i (\pi/2)
J_0 (x)$, and $F_{(1)}$'s are defined by
\beq
F^2_{(1)}&=&1-x_2(1-x_3-x_1).
 \eeq

The threshold resummation form
factor $S_t(x_i)$ is adopted from Ref~.\cite{kurimoto}.It has been
parametrized as \beq
S_t(x)=\frac{2^{1+2c} \Gamma
(3/2+c)}{\sqrt{\pi} \Gamma(1+c)}[x(1-x)]^c,
\label{eq:stxi}
\eeq where the parameter
$c=0.3$. This function is normalized to unity. The evolution factors
$E^{(\prime)}_e$ and $E^{(\prime)}_a$ are given by \beq
E_e(t)&=&\alpha_s(t)\exp[-S_{ab}(t)],\non
E_e^{\prime}(t)&=&\alpha_s(t)\exp[-S_{cd}(t)]|_{b_2=b_1},\non
E_a(t)&=&\alpha_s(t)\exp[-S_{gh}(t)],\non
E_a^{\prime}(t)&=&\alpha_s(t)\exp[-S_{ef}(t)]|_{b_2=b_3},
 \eeq
The Sudakov factors used in the text are defined as
\beq
S_{ab}(t)
&=& s\left(x_1 m_B/\sqrt{2}, b_1\right) +s\left(x_2 m_B/\sqrt{2},
b_2\right) +s\left((1-x_2) m_B/\sqrt{2}, b_2\right) \non
&&-\frac{1}{\beta_1}\left[\ln\frac{\ln(t/\Lambda)}{-\ln(b_1\Lambda)}
+\ln\frac{\ln(t/\Lambda)}{-\ln(b_2\Lambda)}\right],
\label{eq:sab}\\
 S_{cd}(t) &=& s\left(x_1 m_B/\sqrt{2}, b_1\right)
 +s\left(x_2 m_B/\sqrt{2}, b_1\right)
+s\left((1-x_2) m_B/\sqrt{2}, b_1\right) \non
 && +s\left(x_3
m_B/\sqrt{2}, b_3\right) +s\left((1-x_3) m_B/\sqrt{2}, b_3\right)
\non
 & &-\frac{1}{\beta_1}\left[2
\ln\frac{\ln(t/\Lambda)}{-\ln(b_1\Lambda)}
+\ln\frac{\ln(t/\Lambda)}{-\ln(b_3\Lambda)}\right],
\label{Sc}\\
S_{ef}(t) &=& s\left(x_1 m_B/\sqrt{2}, b_1\right)
 +s\left(x_2 m_B/\sqrt{2}, b_2\right)
+s\left((1-x_2) m_B/\sqrt{2}, b_2\right) \non
 && +s\left(x_3
m_B/\sqrt{2}, b_2\right) +s\left((1-x_3) m_B/\sqrt{2}, b_2\right)
\non
 &
&-\frac{1}{\beta_1}\left[\ln\frac{\ln(t/\Lambda)}{-\ln(b_1\Lambda)}
+2\ln\frac{\ln(t/\Lambda)}{-\ln(b_2\Lambda)}\right],
\label{Se}\\
S_{gh}(t) &=& s\left(x_2 m_B/\sqrt{2}, b_2\right)
 +s\left(x_3 m_B/\sqrt{2}, b_3\right)
+s\left((1-x_2) m_B/\sqrt{2}, b_2\right) \non
 &+& s\left((1-x_3)
m_B/\sqrt{2}, b_3\right)
-\frac{1}{\beta_1}\left[\ln\frac{\ln(t/\Lambda)}{-\ln(b_1\Lambda)}
+\ln\frac{\ln(t/\Lambda)}{-\ln(b_2\Lambda)}\right], \label{ww}
\eeq where the function $s(q,b)$ are defined in the Appendix A of
Ref.\cite{luy01}. The scale $t_i$'s in the above equations are
chosen as \beq
t_{a} &=& {\rm max}(\sqrt{x_2} m_B,\sqrt{x_1 x_2}m_B,1/b_1,1/b_2)\;,\non
t_{a}^{\prime} &=& {\rm max}(\sqrt{x_1}m_B,\sqrt{x_1 x_2}m_B,1/b_1,1/b_2)\;,\non
t_{b} &=& {\rm max}(\sqrt{x_2|1-x_3-x_1|}m_B,\sqrt{x_1 x_2}m_B,1/b_1,1/b_3)\;,\non
t_{b}^{\prime} &=& {\rm max}(\sqrt{x_2|x_3-x_1|}m_B,\sqrt{x_1 x_2}m_B,1/b_1,1/b_3)\;,\non
t_{c} &=& {\rm max}(\sqrt{(1-x_2) x_3}m_B, \sqrt{|x_1-x_3|(1-x_2)}m_B,1/b_1,1/b_3)\;,\non
t_{c}^{\prime} &=& {\rm max}(\sqrt{|1-x_2(1-x_3-x_1)|}m_B,
    \sqrt{(1-x_2) x_3} m_B,1/b_1,1/b_3)\;,\non
t_{d} &=&{\rm
max}(\sqrt{(1-x_2)x_3}m_B,\sqrt{(1-x_2)}m_B,1/b_2,1/b_3)\;,
\non
t_{d}^{\prime} &=&{\rm max}(\sqrt{(1-x_2)x_3} m_B,\sqrt{x_3}m_B,1/b_2,1/b_3)\; .
\eeq
\end{appendix}

%%%%%%%%%%%%%%%%%%%%%%%%%%%%%%%%%%%%%%%%%%%%%%%%%%%%%%%%%%%%%%%%%%%%%%%%%%%%%%%%%%%%%%%%%%%%%%5
%                                 reference
%%%%%%%%%%%%%%%%%%%%%%%%%%%%%%%%%%%%%%%%%%%%%%%%%%%%%%%%%%%%%%%%%%%%%%%%%%%%%%%%%%%%%%%%%%%%%%%%%

\end{document}